\documentclass[12pt]{article}
\usepackage{graphicx}
\usepackage{amsmath}
\usepackage{amssymb}

\newcommand{\gs}[1][n]{\ensuremath{g_{#1}^{\star}}}
\newcommand{\qs}[1][n]{\ensuremath{q_{#1}^{\bullet}}}

\newcommand{\qbs}[1][n]{\ensuremath{\bar{q}_{#1}^{\bullet}}}

\newcommand{\qps}[1][n]{\ensuremath{q'{}_{#1}^{\bullet}}}
\newcommand{\qbps}[1][n]{\ensuremath{\bar{q}'{}_{#1}^{\bullet}}}
\newcommand{\qt}[1][n]{\ensuremath{q_{#1}^{\circ}}}
\newcommand{\qbt}[1][n]{\ensuremath{\bar{q}_{#1}^{\circ}}}
\newcommand{\qpt}[1][n]{\ensuremath{q'{}_{#1}^{\circ}}}
\newcommand{\qbpt}[1][n]{\ensuremath{\bar{q}'{}_{#1}^{\circ}}}
\newcommand{\qb}{\ensuremath{\bar{q}}}
\newcommand{\g}{\ensuremath{g_{{}_5}}}

\newcommand{\shat}{\ensuremath{\hat{s}}}
\newcommand{\that}{\ensuremath{\hat{t}}}
\newcommand{\uhat}{\ensuremath{\hat{u}}}
\newcommand{\tn}{\ensuremath{\hat{t}{'}}}
\newcommand{\un}{\ensuremath{\hat{u}{'}}}
\begin{document}

\begin{titlepage}
\begin{flushright}
OSU-HEP-02-01\\
\end{flushright} \vskip 2cm
\begin{center}
{\Large\bf Collider Implications of Universal Extra Dimensions}
\vskip 1cm {\normalsize\bf
C.\ Macesanu,\footnote{email: mcos@pas.rochester.edu} C.D.\
McMullen\footnote{email: mcmulle@okstate.edu} and S.\
Nandi\footnote{email:
shaown@okstate.edu}\addtocounter{footnote}{-3}
\\} \vskip 0.5cm
{\it Department of Physics, Oklahoma State University\\
Stillwater, OK~~74078, USA\\[0.1truecm]
}

\end{center}
\vskip 2.5cm

\begin{abstract}

We consider the universal extra dimensions scenario of Appelquist,
Cheng, and Dobrescu, in which all of the SM fields propagate into
one extra compact dimension, estimated therein to be as large as
$\sim (350$ GeV$)^{-1}$. Tree-level KK number conservation
dictates that the associated KK excitations can not be singly
produced. We calculate the cross sections for the direct
production of KK excitations of the gluon, $\gs$, and two distinct
towers of quarks, \qs\ and $\qt$, in proton-antiproton collisions
at the Tevatron Run I and II energies in addition to proton-proton
collisions at the Large Hadron Collider energy. The experimental
signatures for these processes depend on the stability of the
lowest-lying KK excitations of the gluons and light quarks. We
find that the Tevatron Run I mass bound for KK quark and gluon
final states is about $350$--$400$ GeV, while Run II can push this
up to $450$--$500$ GeV at its initial luminosity and $500$--$550$
GeV if the projected final luminosity is reached. The LHC can
probe much further: The LHC will either discover UED KK
excitations of quarks and gluons or extend the mass limit to about
$3$ TeV.

\end{abstract}

\end{titlepage}

\newpage

\noindent
{\bf 1.  Introduction}

\vspace{0.2cm}

\noindent The low-energy phenomenology of superstring-inspired
models with large extra compact dimensions depends on the
mechanism of new physics by which the Standard Model (SM) fields
are constrained, if at all, to motion in the usual $3$D wall
(D$_3$ brane) of the usual three spatial dimensions. It might
naively be speculated that as more SM fields are free to propagate
into the extra compact dimensions (the bulk), then the collider
bounds on the compactification scale would significantly
strengthen. A non-universal model where the gauge bosons propagate
into the bulk, but the fermions are confined to the usual SM D$_3$
brane, for example, does produce more stringent collider bounds
than a model where all of the SM fields are confined to the D$_3$
brane. However, scenarios with universal extra dimensions (UED),
in which all of the SM fields propagate into the bulk, have much
weaker collider bounds. This is due to tree-level Kaluza-Klein
(KK) number conservation, which dictates that colliding SM initial
states can not produce single KK excitations and also forbids
tree-level indirect collider effects. In the non-universal
scenarios, the SM fields that are confined to the D$_3$ brane
appear in the Lagrangian with delta functions, thereby permitting
couplings that violate KK number conservation.

Only the gravitons propagate into the extra compact dimensions in
the class of models based on the approach of Arkani-Hamed,
Dimopoulos, and Dvali (ADD)~\cite{Planck}, where the
compactification is symmetric -- \textit{i.e.}, all of the $N$
extra dimensions have the same compactification radius $R$. The
string scale $M_D$ is much smaller than the four-dimensional
Planck scale $M_P$ \cite{superstring}, which are related by
$M_{P}^2 = M_D^{N+2} R^N$. Any SM fields that propagate into the
bulk would have (KK) excitations with masses at the $10$ MeV scale
or less. The non-observation of such states up to about a TeV
implies, in this class of models, that all of the SM fields are
confined to the usual SM D$_3$ brane. Hence, the only source of
new contributions to collider processes arises from the KK
excitations of the graviton. Although the contributions of
individual KK modes, with $4$D gravitational strength, to collider
processes is extremely small, a very large number of such modes
contribute in a TeV-scale collider process because the
compactification scale {$\mu$} is so small ($\mu \sim $ mm$^{-1}
\sim 10^{-3}$ eV). The net KK effect can cause a significant
deviation from the SM production rates. Bounds on the string scale
from analyses of various collider processes are typically on the
order of a TeV~\cite{collider,HLZ} for these symmetric
compactification models.

One way to permit some or all of the SM fields to propagate into
the bulk is to relax the constraint that the extra compact
dimensions be symmetric. Let us first consider the case where only
the SM gauge bosons propagate into the bulk.  As an example, it is
possible to devise a model with asymmetrical compactification with
five TeV$^{-1}$-size extra compact dimensions and one mm-size
extra dimension, where the SM gauge bosons (and perhaps the Higgs
boson) propagate into one of the TeV$^{-1}$-size dimensions. It
was shown in Ref.~\cite{asym} that this model satisfies all of the
current astrophysical and cosmological constraints~\cite{astro}.
These asymmetric scenarios have a more direct effect in
high-energy collider processes. Originating with the suggestion by
Antoniadis~\cite{Antoniadis}, some of the studies that have been
done for the collider phenomenology of the scenario in which the
SM gauge bosons can propagate into the bulk, but where the SM
fermions can not~\cite{asymcoll}, include: the effects on
electroweak (EW) precision measurements~\cite{ew}, Drell-Yan
processes in hadronic colliders~\cite{muon}, $\mu^{+} \mu^{-}$
pair production in electron-positron colliders~\cite{muon}, EW
processes in very high-energy electron-positron
colliders~\cite{estar}, and multijet production in very
high-energy hadronic colliders~\cite{gstar}. The typical bound on
the compactification scale is $1$--$2$ TeV.

The UED model, where all of the SM fields propagate into one or
more extra compact dimensions, may intuitively seem more natural
than selectively confining SM fields to the usual SM D$_3$ brane.
This scenario may be thought of as a generalization of the usual
SM wall to a D$_{3+N}$ brane, where $N$ represents the number of
extra compact dimensions into which the SM fields propagate. In
this universal model of Appelquist, Cheng, and Dobrescu
\cite{ACD}, KK number conservation governs all of the couplings
involving KK excitations. In particular, each such vertex involves
at least two KK excitations. At the tree-level, then, KK effects
can not manifest themselves indirectly at colliders, and direct
production is only possible in pairs of KK states. Although KK
number conservation is broken at the one-loop level, the
lowest-lying KK excitations of the light fermions and the massless
gauge bosons do not decay to the SM zero-modes at any order
without a special mechanism to support this decay. Thus, the
lowest-lying KK excitations of the light fermions and the massless
gauge bosons may be completely stable. Possible decay mechanisms
have been proposed in the literature~\cite{ACD,Rujula,Rizzo}.
Collider bounds for this universal scenario are comparatively
light: The current mass bound \cite{ACD,Rizzo,UED} for the first
KK excited modes is relatively low ($\sim 350$-$400$ GeV).

In this work, we make a detailed study of the collider
implications of the universal scenario, in which all of the SM
fields propagate into one TeV$^{-1}$-size extra compact dimension.
More specifically, we calculate the cross sections for the
pair-production of KK excitations of the gluons, $\gs$, and two
distinct KK quark towers, \qs\ and $\qt$, in proton-antiproton
collisions at the Tevatron Run I and II energy in addition to
proton-proton collisions at the Large Hadron Collider (LHC)
energy. The signatures of these KK excitations depend on the
stability of the lowest-lying KK excitations of the light quarks
and gluons. We find that the Tevatron Run I mass bound for KK
quark and gluon final states is about $350$--$400$ GeV, while Run
II can push this limit up to $450$--$550$ GeV, depending on the
luminosity. The LHC can probe much further: The LHC will either
discover UED KK excitations of the quarks and gluons or extend the
mass limit to about $3$ TeV. The organization of our paper is as
follows. We develop the key ingredients of our formalism in
Section $2$, which is supplemented by additional details in the
Appendix. We also present the Feynman rules involving the KK
excitations of the gluons and quarks. Section $3$ contains our
analytical expressions for the pair-production of KK excitations
of the gluons and quarks. We treat the case of stable KK final
states in Section $4$. Here we present our results for the
production cross sections of pairs of stable KK excitations, and
discuss how to search for their collider signatures. We discuss
possible decay mechanisms in Section $5$. Our results for the case
where the pair-produced KK final states decay may be found here,
along with methods of searching for this associated collider
phenomenology. We present our conclusions in Section $6$.

\vspace{0.5cm}

\noindent
{\bf 2.  Formalism}

\vspace{0.2cm}

\noindent We are interested in the collider implications of the
universal scenario, in which all of the SM fields propagate into a
single TeV$^{-1}$-size extra compact dimension. Our focus is on
the tree-level parton subprocesses that involve the direct
pair-production of KK excitations of gluons, $\gs$, and two
distinct KK quark towers, \qs\ and $\qt$. We begin by generalizing
the usual $4$D Lagrangian density to its $5$D analog. We perform
orbifold compactification and integrate over the fifth dimension
$y$ to obtain the effective $4$D theory, which is the usual $4$D
Lagrangian density plus new physics terms involving the KK
excitations of the quark and gluon fields. These new terms provide
the masses of the KK modes as well as the Feynman rules for the
vertices and propagators involving KK excitations. We develop the
key elements of our formalism here, while supplementary details
are included in the Appendix.

We denote the $4$D SM quark multiplets for one generation by
$Q_L^{\textnormal{SM}}(x)$, $U_R^{\textnormal{SM}}(x)$, and
$D_R^{\textnormal{SM}}(x)$. For example, the first generation is:

\vspace{-9pt} \begin{equation} Q_L^{\textnormal{SM}}(x) = q_L(x) =
\left( \! \begin{array}{c} u (x)
\\ d (x) \end{array} \! \right)_{\!L} \, , \;\;
U_R^{\textnormal{SM}}(x) = u_R(x) \, , \;\;
D_R^{\textnormal{SM}}(x) = d_R(x) \, . \end{equation}

\noindent Each $4$D state is a two-component Weyl spinor. The
analogous $5$D quark multiplets consist of massless four-component
vector-like quarks, which we denote by $Q(x,y)$, $U(x,y)$, and
$D(x,y)$. When these $5$D fields are decomposed into $4$D fields,
corresponding to each $4$D field are a left-handed and
right-handed zero mode. Each mode is a two-component Weyl spinor
in $4$ dimensions. Half of the zero modes, which are not present
in the $4$D SM, may be projected out via the simple orbifold
compactification choice, $S_1 / Z_2$ $( Z_2$:$y\rightarrow -y )$.
The gauge fields polarized along the usual SM directions must be
even under $y \rightarrow -y$ such that the zero modes will
correspond to the usual $4$D gauge fields, which implies that the
gauge fields polarized along the $y$ direction must be odd. For
the quark fields, each of the KK $(n > 0)$ modes for each
multiplet will have a left-chiral and right-chiral part. The
$Q_L^n (x)$, $U_R^n (x)$, and $D_R^n (x)$ components must be
associated with the part of $Q(x,y)$, $U(x,y)$, and $D(x,y)$ that
is even under $y \rightarrow -y$ in order to recover the
appropriate SM chiral zero mode states. The remaining components,
$Q_R^n (x)$, $U_L^n (x)$, and $D_L^n (x)$, must be associated with
the part of $Q(x,y)$, $U(x,y)$, and $D(x,y)$ that is odd under $y
\rightarrow -y$ such that the zero modes not observed in the SM
will be projected out. Each of the $5$D multiplets $Q(x,y)$,
$U(x,y)$, and $D(x,y)$ can therefore be Fourier expanded in terms
of the compactified dimension $y$ as

\vspace{-9pt} \begin{eqnarray} Q (x,y) &\!\!\! = &\!\!\!
\frac{1}{\sqrt{\pi R}} \left\{ \left( \! \begin{array}{c} u (x)
\\ d (x) \end{array} \! \right)_{\!L} + \sqrt{2} \sum_{n=1}^{\infty}
\left[ Q_L^n (x) \cos \left(\frac{n y}{R} \right) + Q_R^n (x) \sin
\left(\frac{n y}{R} \right) \right] \right\} \\
U (x,y) &\!\!\! = &\!\!\! \frac{1}{\sqrt{\pi R}} \left\{ u_R (x) +
\sqrt{2} \sum_{n=1}^{\infty} \left[ U_R^n (x) \cos \left(\frac{n
y}{R} \right) + U_L^n (x) \sin \left(\frac{n y}{R} \right) \right]
\right\} \\
D (x,y) &\!\!\! = &\!\!\! \frac{1}{\sqrt{\pi R}} \left\{ d_R (x) +
\sqrt{2} \sum_{n=1}^{\infty} \left[ D_R^n (x) \cos \left(\frac{n
y}{R} \right) + D_L^n (x) \sin \left(\frac{n y}{R} \right) \right]
\right\} \, .
\end{eqnarray}

The SM fermion masses arise from the Yukawa couplings through the
Higgs vacuum expectation value (VEV), while the KK modes receive
mass from the kinetic term in the $5$D Lagrangian density as well
as from the Yukawa couplings via the Higgs VEV's. We first
calculate the mass arising from the kinetic term. The $5$D
Lagrangian density for the kinetic terms and interactions of the
$5$D gluon field $A_M^a(x,y)$ with the $5$D $Q(x,y)$ fields are:

\vspace{-9pt} \begin{equation} \label{eq:LkinQ} \mathcal{L}_5  = i
\bar{Q} (x,y) \left\{ \Gamma^M \left[ \partial_M + i \g T^a A_M^a
(x,y) \right] \right\} Q (x,y) \, .
\end{equation}

\noindent Here, \g\ is the $5$D strong coupling, $M$ is the $5$D
analog of the Lorentz index $\mu$, \textit{i.e.}, $M \in
\{\mu,4\}$, and the $5$D gluon fields $A_M^a (x,y)$ can be Fourier
expanded in terms of the compactified extra dimension $y$ as:

\vspace{-9pt} \begin{eqnarray} A_{\mu}^a (x,y) =
\frac{1}{\sqrt{\pi R}}\left[ A_{\mu 0}^a (x) +
\sqrt{2} \sum_{n=1}^{\infty}A_{\mu ,n}^{a} (x) \cos(\frac{n y}{R}) \right] \\
A_{4}^a (x,y) = \frac{\sqrt{2}}{\sqrt{\pi R}}
\sum_{n=1}^{\infty}A_{4,n}^{a} (x) \sin(\frac{n y}{R}) \, .
\end{eqnarray}

\noindent The normalization of $A_0^a (x)$ is one-half that of the
$n > 0$ modes, necessary to obtain canonically normalized kinetic
energy terms for the gluon fields in the effective $4$D Lagrangian
density~\cite{rescale}. As previously stated, under the
transformation $y\rightarrow -y$, the decomposed gluon fields
transform as $A_{\mu}^a (x,-y) = A_{\mu}^a (x,y)$ and $A_4^a
(x,-y) = - A_4^a (x,y)$. We choose to work in the unitary gauge,
where we can apply the gauge choice $A_{4,n}^a (x) = 0$
\cite{gauge}.

Integrating the kinetic part of Eq.~(\ref{eq:LkinQ}) over the
compactified dimension $y$ yields the $4$D Lagrangian density, and
similarly for $U (x,y)$ and $D (x,y)$. This effective $4$D
Lagrangian density consists of the usual kinetic terms for the SM
fields, kinetic terms for the massive Dirac spinors $Q^n (x)$,
$U^n (x)$, and $D^n (x)$, and mass terms for the KK excitations
with mass $M_n^{\mathit{KK}} = n/R = n \mu$, where $\mu$ is the
compactification scale, $1/R$.

Thus, in the absence of the Higgs mechanism, the KK excitations
have masses given by $M_n = M_n^{\mathit{KK}} = n/R = n \mu$.
Additional mass contributions from the Yukawa couplings of the
$5$D quark multiplets via the Higgs VEV's are obtained by writing
the $5$D Lagrangian density for the couplings of the $5$D quark
multiplets to the $5$D Higgs field, Fourier expanding these $5$D
fields in terms of the compactified dimension $y$, and integrating
over the extra dimension. The eigenvalues of the resulting mass
matrix give the net mass $M_n$ of the KK modes in terms of the
mass of the corresponding quark field $M_q$ and the mass from the
compactification $M_n^{\mathit{KK}}$:

\vspace{-9pt} \begin{equation} \label{eq:mass} M_n =
\sqrt{(M_n^{\mathit{KK}})^2 + M_q^2} \, .
\end{equation}

\noindent Relative to the compactification scale, the SM quark
masses are negligible except for the top mass $M_t$.

The QCD interactions involving KK excitations include purely
gluonic couplings as well as couplings with quark fields. The
purely gluonic case was discussed in detail in Ref.~\cite{gstar},
and the resulting couplings are identical to those of this
universal scenario. We therefore refer the reader to this prior
work for these details, and concentrate on the couplings of quark
fields to gluon fields. The Feynman rules for the QCD interactions
involving the KK excitations of the gluons and the two towers of
KK excitations corresponding to each of the quark fields can be
obtained by integrating the second part of Eq.~(\ref{eq:LkinQ})
over the compactified dimension $y$ via Fourier expansion of the
$5$D fields in terms of $y$, and similarly for $U (x,y)$ and $D
(x,y)$.

Each KK \qs\ and \qt\ state is identified as a combination of $Q$,
$U$, and $D$. In the limit of massless SM quarks, this combination
can be expressed as:

\vspace{-9pt} \begin{equation} \label{eq:id} Q_{L,R}^n (x) \equiv
P_{L,R} \left( \! \begin{array}{c} u_n^{\bullet} (x)
\\ d_n^{\bullet} (x) \end{array} \! \right) \, , \;\;
U_{R,L}^n (x) \equiv P_{R,L} u_n^{\circ} \, , \;\; D_{R,L}^n (x)
\equiv P_{R,L} d_n^{\circ} \, ,
\end{equation}

\noindent where the projection operators are defined as $P_{L,R}
\equiv \frac{1}{2} (1 \mp \gamma_5 )$. In general, there is an
additional Yukawa contribution to the masses, in which the $U_R$
and $U_L$ fields contribute to the mass of the \qs\ via the Higgs
VEV, and similarly for contributions to \qt\ from $Q_L$ and $Q_R$.
For example, taking the SM $c$ quark to be massless, the
combination of the second-generation up-type quark component of
the KK multiplet $Q_{2L}^n(x)$ with the second-generation up-type
quark component of $Q_{2R}^n(x)$ is identified as the single KK
charm quark $c^{\bullet}$, which receives KK mass $M_n = n \mu =
1/R$ from the kinetic term. There is a second KK tower
corresponding to the SM charm quark, which comes from
$U_{2R}^n(x)$ and $U_{2L}^n(x)$, that we denote by $c^{\circ}$. By
\gs\ we denote KK mode $n$ of the gluon, and by \qs\ and \qt\ we
denote KK mode $n$ of two distinct towers of KK excitations of a
given SM quark field $q$. Each KK quark tower contains terms that
are even and odd under $Z_2$ parity. However, in KK quark pair
production, the KK final states will be polarized with helicity
corresponding to their even states ($Q_L(x)$, $U_R(x)$, and
$D_R(x)$)in the cross channels, and the components associated with
the odd part of the $5$D fields ($Q_R(x)$, $U_L(x)$, and $D_L(x)$)
will only show up in direct channel production.\footnote{This
relies on the expansion in Eq.~\ref{eq:id}, which is valid for KK
excitations of massless SM quarks. Massive KK quarks receive an
additional small mass contribution from the Higgs mechanism. Also,
recall that we are working in the unitary gauge with gauge choice,
$A_4^{an}(x) = 0$.} For KK quark-gluon production, the final KK
states will again be polarized with helicity corresponding to the
even states. This is because the projection operators ensure the
conservation of $Z_2$ parity. Regarding our notation, $n$ will be
strictly nonzero unless we explicitly state otherwise.

The detailed procedure for integrating over the fifth dimension
$y$ to obtain, in the effective $4$D theory, the factors for the
allowed vertices involving the \qs\ and \qt\ fields may be found
in the Appendix, and lead to the coupling strengths displayed in
Fig.~\ref{fig:feynq}. The states with helicity corresponding to
the odd states under $Z_2$ parity ($Q_R(x)$, $U_L(x)$, and
$D_L(x)$) only appear in couplings involving \qs\ \qs\ or \qt
$\qt$, and do not show up when a SM quark is present. A SM quark
can only couple to KK states with helicity corresponding to the
even states (($Q_L(x)$, $U_R(x)$, and $D_R(x)$). The triple KK
vertices with \qs\ and \qt\ fields involve the integration of
three cosines for the even parts and one cosine and two sines for
the odd parts. This latter integration results in a minus sign
relative to the first one whenever the KK gluon is more massive
than either KK quark, which results in the presence of a
$\gamma_5$ in these vertices. Note also that the two towers \qs\
and \qt\ do not couple to one another. The Feynman rules for the
purely gluonic vertices are summarized in Ref.~\cite{gstar}.
Notice that a single KK mode can not couple to SM fields. This is
a consequence of the more general tree-level conservation of KK
number, which dictates that $N$ KK modes,
$n_{{}_{1}}$,$n_{{}_{2}}$,$\ldots$,$n_{{}_N}$, can only couple to
one another if they satisfy the relation:


\begin{figure}
\setlength{\abovecaptionskip}{10pt}
\centering{\includegraphics[bb=72 180 513 684]{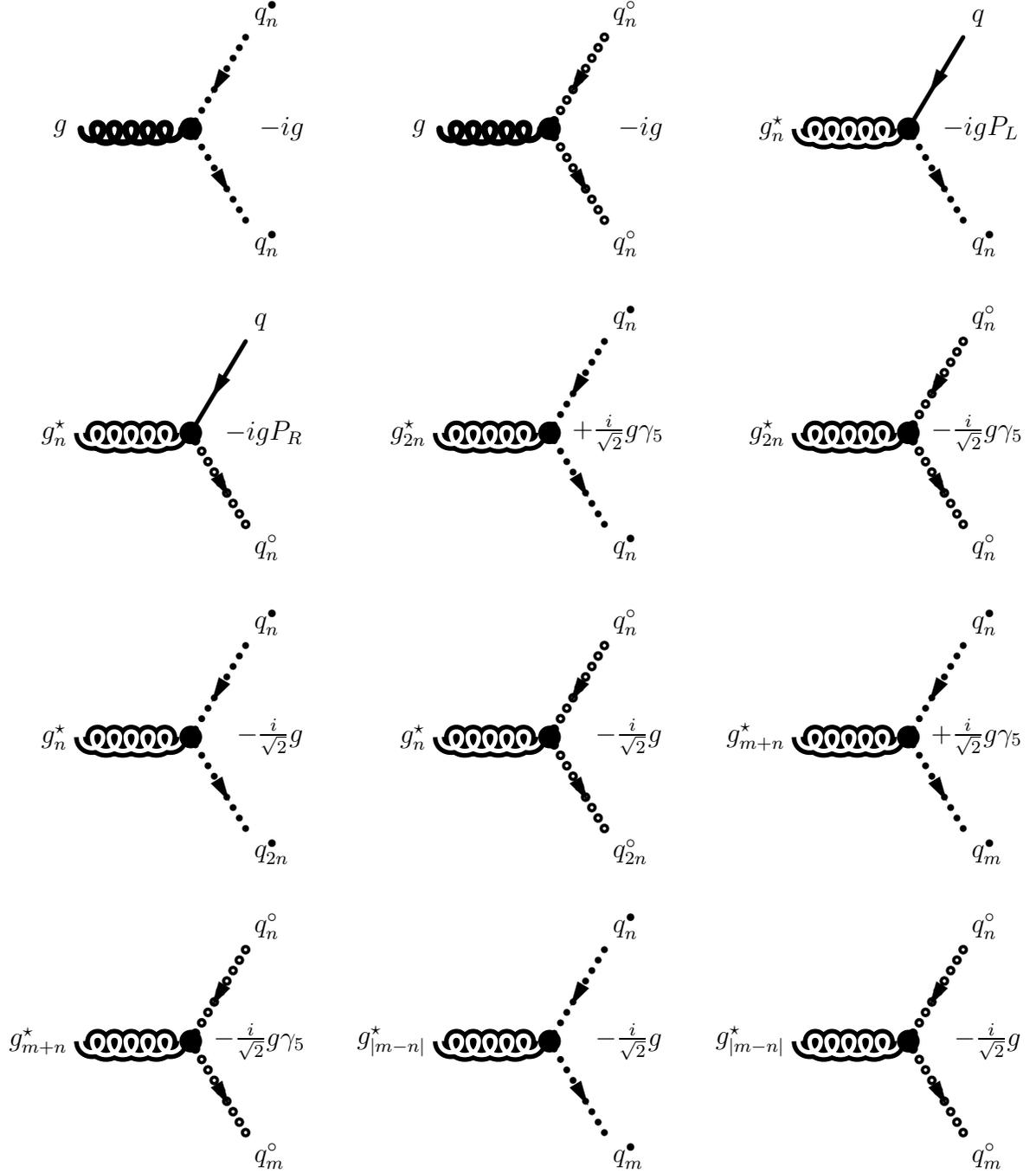}}
\caption{\small{Relative coupling strengths of vertices involving
$\qs$'s and $\qt$'s. Only the overall factors are shown:  These
vertices also involve the usual SU($3$) matrix element and the
Dirac $\gamma_{\mu}$ matrix. Here, $n$ and $m$ are distinct
positive integers ($n  \neq m$) and the projection operators are
defined as $P_{L,R}\equiv(1 \mp \gamma_5)/2$.}} \label{fig:feynq}
\setlength{\abovecaptionskip}{10pt}
\end{figure}

\vspace{-9pt} \begin{equation} \label{eq:modes} \mid \! n_{{}_{1}}
\, \pm \, n_{{}_{2}} \, \pm \, \cdots \, \pm \, n_{{}_{N-1}} \!
\mid \,= n_{{}_N} \, .
\end{equation}

\noindent KK number conservation strictly applies at every vertex,
as well as for tree-level $N\rightarrow M$ processes, but is
broken at the loop-level. The higher modes can therefore decay to
the lower modes at the loop-level, but the lowest-lying KK modes
of the light quarks and massless gluons will be completely stable
unless there exists another form of new physics to serve as a
decay mechanism. We will return to this point in Section~$5$.

The \gs\ propagator is that of a usual massive gauge boson, shown
here in the unitary gauge:

\vspace{-9pt} \begin{equation} \label{eq:prop} - i \Delta_{\mu\nu
n}^{ab}(p^2) = -i \delta^{ab} \frac{g_{\mu \nu}  - \frac{p_{\mu}
p_{\nu}}{M_n^2}}{p^2  -  M_n^2  +  i M_n \Gamma_g^n} \, .
\end{equation}

\noindent Similarly, the \qs\ and \qt\ propagators have the form
of a usual massive quark:

\vspace{-9pt} \begin{equation} \label{eq:propq} - i
\Delta_n^{a'b'}(p^2) = i \delta^{a'b'} \frac{\not \!p + M_n}{p^2 -
M_n^2 + i M_n \Gamma_q^n} \, .
\end{equation}

\noindent The decay widths of the $\gs$'s, $\qs$'s, and $\qt$'s
depend on stability of the lowest-lying KK excitations of the up
quark, down quark, and gluon. However, these decay widths are
immaterial for production processes, since KK number conservation
forbids any $s$-channel KK propagators from arising in tree-level
subprocesses with initial SM fields.

The mass of the \gs\ also enters into the expression for the cross
section via summations over polarization states when external
$\gs$'s are involved. For the direct production of a $\gs$, the
summation of polarization states is given by

\vspace{-9pt} \begin{equation} \sum_{\sigma} \epsilon_{\mu
n}^{a\ast}(k,\sigma)\epsilon_{\nu n}^b(k,\sigma)  =
\mbox{\raisebox{-.6ex}{\huge $($}} \!\! -g_{\mu \nu}
 +  \frac{k_{\mu}k_{\nu}}{M_n^{2}}\mbox{\raisebox{-.6ex}{\huge
$)$}}\delta^{ab} \, .
\end{equation}

\noindent For the case of external $g$'s, a projection such as

\vspace{-9pt} \begin{equation} \sum_{\sigma}
\epsilon_{\mu}^{a\ast}(k,\sigma)\epsilon_{\nu}^b(k,\sigma)  =
\mbox{\raisebox{-.6ex}{\huge $[$}} \! -g_{\mu \nu} +
\frac{(\eta_{\mu}k_{\nu}  +  \eta_{\nu}k_{\mu})}{(\eta\cdot k)} -
\frac{\eta^2 k_\mu k_\nu}{(\eta\cdot
k)^2}\mbox{\raisebox{-.6ex}{\huge $]$}}\delta^{ab}
\end{equation}

\noindent can be made to eliminate unphysical longitudinal
polarization states (and thereby satisfy gauge invariance), where
$\eta_{\mu}$ is an arbitrary four-vector.

\vspace{0.5cm}

\noindent {\bf 3.  Pair Production of KK Excitations}

\vspace{0.2cm}

\noindent We have in mind the production of pairs of KK
excitations of the gluons, $\gs$, and quarks, \qs\ and $\qt$, in
proton-antiproton collisions at the Tevatron Run I or II energy or
proton-proton collisions at the LHC energy. We focus on the parton
subprocesses in this section and postpone numerical results to the
following sections where the stability of the lowest-lying KK
excitations is addressed. The various subprocesses are enumerated
in Table~\ref{table:process}. We perform our calculations at the
tree-level, and restrict ourselves to two final states. Due to KK
number conservation, not only must the KK excitations be produced
in pairs, but they necessarily have the same mode $n$, which is
the same mode that any KK propagators will have. We neglect the
quark masses except for the top mass $M_t$, but neglect the
content of top flavor in the colliding protons and antiprotons.
Thus, the top quark only enters into the calculation of the cross
sections for $g g \rightarrow \qs \qbs$ and $q \bar{q} \rightarrow
\qps \qbps$, and the analogous subprocesses for the $\qt$'s. We
also neglect the decay widths of all SM and KK particles in this
section since massive propagators will not appear in the
$s$-channel due to tree-level KK number conservation and our
neglect of initial top quarks. We will incorporate the decay
widths in the subsequent decay of the final states in Section $5$,
where we discuss possible mechanisms for the decay of the
lowest-lying KK states.
\begin{table}
\centering{\includegraphics[height = 1.8in,width=5.in]{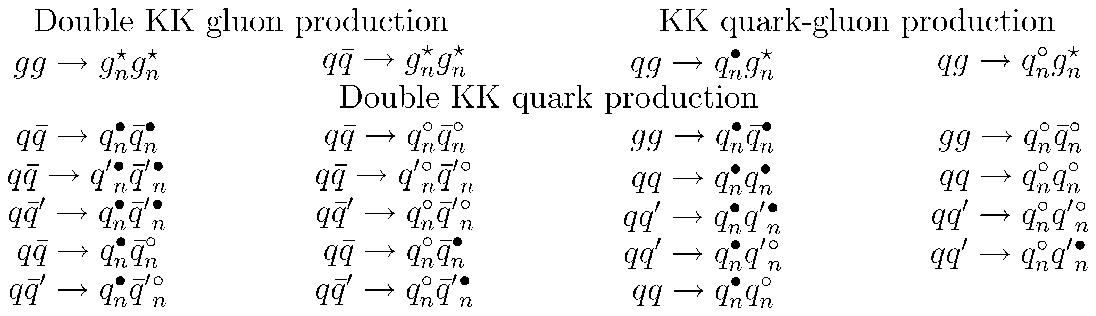}}
\caption{\small{Subprocesses leading to double KK production at
hadronic colliders. Not shown are subprocesses that are simply
related by the exchange of a particle and antiparticle, as in $\qb
g \rightarrow \qbs \gs$.}} \label{table:process}
\end{table}

Double KK gluon production subprocesses consist of $g g
\rightarrow \gs \gs$ and $q \bar{q} \rightarrow \gs \gs$. The
former subprocess involves direct-channel SM gluon exchange,
cross-channel KK gluon exchanges, and the four-point interaction.
The latter subprocess is unique in that there are five tree-level
Feynman diagrams, which include direct-channel SM gluon exchange
and cross-channel \qs\ and \qt\ exchanges. For the purely gluonic
subprocess, the
amplitude-squared,\addtocounter{footnote}{-1}\footnote{We employ
\textit{FORM}~\cite{FORM}, a symbolic manipulation program, in the
evaluation of the squares of the amplitudes. The expressions in Eqs. 
(\ref{e1})--(\ref{e9}) agree with the results of \cite{Webber}.} summed over final
states and averaged over initial states, is:

\vspace{-8pt} \begin{eqnarray}\label{e1}
\mbox{\raisebox{-.5ex}{\Large$\bar{\Sigma}$}}
 \left| {\cal M}(g g \rightarrow \gs \gs ) \right| & =  &
{ 9  \over 4}\ g_s^4(Q) \left(1 - {\tn \un \over \shat^2}  \right)
\nonumber \\
&\!\!\! \cdot &\!\!\!\!\!\! \left[ 3 \ +\ 2 \frac{\shat^2}{\tn^2 \un^2}
\left(3 M_n^4 +\shat^2 \right) \ - \ 2 \frac{\shat}{\tn \un}
\left(3 M_n^2 +2 \shat \right)\right] \, ,
\end{eqnarray}

\noindent where the strong 
coupling constant $g_s$ is evaluated at the scale $Q$ 
equal to the mass of the
final state KK excitations $M_n$, and $ \hat{v} '$ represents
subtraction of $M_n^2$ from the Mandelstam variable $ \hat{v} \in
\{ \shat , \that , \uhat \}$ (\textit{i.e.}, $ \hat{v} ' =
\hat{v} - M_n^2$). We note that $g g \rightarrow \gs \gs$ is the
same in the UED scenario considered here as well as in a model
where only gluons propagate into the bulk. However, each of the
remaining subprocesses is different. The amplitude-squared for $q
\bar{q} \rightarrow \gs \gs$ is:

\vspace{-8pt} \begin{eqnarray}
\mbox{\raisebox{-.5ex}{\Large$\bar{\Sigma}$}}\! \mid \!
\mathcal{M}(q \bar{q}  \rightarrow \gs \gs )\!\mid ^2 \, =&
\!\!\!& \!\!\!\!\!\!\!\!\! \frac{2}{27}\ g_s^4(Q)\left[
\frac{M_n^2}{\shat} \left( -4 \frac{\shat^4}{\tn^2 \un^2} + 57
\frac{\shat^2}{\tn \un} - 108 \right) \right.
\nonumber \\
&\!\!\! + &\!\!\!\! \left. 20 \frac{\shat^2}{\tn \un} - 93 + 108
\frac{\tn \un}{\shat^2} \right] \, .
\end{eqnarray}

KK quark-gluon production results from $q g \rightarrow \qs \gs$
and $q g \rightarrow \qt \gs$. (We will not enumerate subprocesses
that are simply related by particle-antiparticle replacement, such
as $\bar{q} g \rightarrow \qbs \gs$.) These subprocesses involve
$s$-channel SM quark exchange, $t$-channel \gs\ exchange, and
$u$-channel KK quark exchange. The square of the matrix element
for $q g \rightarrow \qs \gs$ is:

\vspace{-8pt} \begin{eqnarray}
\mbox{\raisebox{-.5ex}{\Large$\bar{\Sigma}$}}\! \mid \!
\mathcal{M}(q g  \rightarrow \qs \gs)\!\mid ^2 \, =& &
\!\!\!\!\!\!\!\!\! \frac{-1}{3}\ g_s^4(Q)\left( 
\frac{5 \shat^2}{12 \tn^2} +  \frac{\shat^3}{\tn^2 \un} +
\frac{11 \shat \un}{6 \tn^2} + \frac{5\un^2}{12\tn^2} +
\frac{\un^3}{\shat \tn^2}\right) \, .
\end{eqnarray}

\noindent The subprocess $q g \rightarrow \qt \gs$ is identical to
$q g \rightarrow \qs \gs$. That is, the sign of the $\gamma_5$
matrix is not important in KK quark production unless both \qs\
and \qt\ are involved in the same subprocess, \textit{e.g.}, in $q
\bar{q} \rightarrow \gs \gs$ or $q q \rightarrow \qs \qt$.

Subprocesses with identical final \qs\ or \qt\ states feature $t$-
and $u$-channel \gs\ exchanges. A relative minus sign represents
the antisymmetrization of fermionic wave functions that originates
from the interchange of identical fermionic states between the two
diagrams. Notice that although a given SM quark $q$ and its KK
counterparts have different mass, they have the same fermionic
properties that produces the minus sign for the antisymmetrization
of wave functions. The amplitude-squared for $\qs \qs$ production
is:

\vspace{-8pt} \begin{eqnarray}
\mbox{\raisebox{-.5ex}{\Large$\bar{\Sigma}$}}\! \mid \!
\mathcal{M}(q q  \rightarrow \qs \qs)\!\mid ^2 \, =& \!\!\!&
\!\!\!\!\!\!\!\!\! \frac{1}{27}\  g_s^4(Q)\left[
\frac{M_n^2}{\shat} \left( -6 \frac{\shat^4}{\tn^2\un^2} + 17
\frac{\shat^2}{\tn\un} \right) \right.
\nonumber \\
&\!\!\! + &\!\!\!\! \left. 6 \frac{\shat^4}{\tn^2\un^2} -16
\frac{\shat^2}{\tn\un} + 2 \right] \, .
\end{eqnarray}

\noindent The identical result is obtained for $\qt \qt$
production.

Double KK quark-antiquark pairs with the same flavor can arise
from initial gluons or quarks. The former case involves
direct-channel SM gluon exchange and cross-channel KK quark
exchanges. The latter case consists of $s$-channel SM gluon
exchange, and, in the case of initial partons of the same flavor
as the final states, $t$-channel \gs\ exchange. For initial
gluons, squaring the amplitude leads to the following expression
for KK quark pair production:

\vspace{-8pt} \begin{eqnarray}
\mbox{\raisebox{-.5ex}{\Large$\bar{\Sigma}$}}\! \mid \!
\mathcal{M}(g g  \rightarrow \qs \qbs)\!\mid ^2 \, =& \!\!\!&
\!\!\!\!\!\!\!\!\!  g_s^4(Q)\left[
\left( \frac{M_n^4}{\tn \un}-\frac{M_n^2}{\shat} \right) \left( \frac{3}{2} -
\frac{2 \shat^2}{3 \tn\un} \right) \right. \nonumber \\
&\!\!\! + &\!\!\!\! \left. 
\frac{\shat^2}{6 \tn\un} - \frac{17}{24} + \frac{3 \tn\un}{4 \shat^2}
 \right] \, ,
\end{eqnarray}

\noindent where the only difference for the case of KK top pair
production is adjustment of the mass via Eq.~\ref{eq:mass}. The
amplitude-squared for KK quark-antiquark final states arising from
SM quark-antiquark initial states, for which the flavor is the
same in the initial and final states, is:

\vspace{-8pt} \begin{eqnarray} \label{eq:dqqbqqb}
\mbox{\raisebox{-.5ex}{\Large$\bar{\Sigma}$}}\! \mid \!
\mathcal{M}(q \bar{q}  \rightarrow \qs \qbs)\!\mid ^2 \, =&
\!\!\!& \!\!\!\!\!\!\!\!\! \frac{1}{54}\ g_S^4(Q)\left[
\frac{M_n^2}{\shat} \left( 48 - 12 \frac{\shat}{\tn} + 12
\frac{\shat^2}{\tn^2} \right) \right.
\nonumber \\
&\!\!\! + &\!\!\!\! \left. 48 \frac{\tn^2}{\shat^2} + 36
\frac{\tn}{\shat} + 23 + 16 \frac{\shat}{\tn} + 12
\frac{\shat^2}{\tn^2} \right] \, .
\end{eqnarray}

\noindent This does not lead to KK top quark production since the
top quark content of the colliding protons is negligible. The
relative sign between the two diagrams again incorporates the
antisymmetrization of fermionic wave functions corresponding to
the interchange of two fermionic states between the two diagrams.
When the final states have different flavors than the initial
state, only the $s$-channel contributes. For the lighter flavors,
this is simply the $s$-channel part of Eq.~\ref{eq:dqqbqqb}:

\vspace{-8pt} \begin{eqnarray} \label{eq:dqqbqpqpb}
\mbox{\raisebox{-.5ex}{\Large$\bar{\Sigma}$}}\! \mid \!
\mathcal{M}(q \bar{q}  \rightarrow \qps \qbps)\!\mid ^2 \, =& &
\!\!\!\!\!\!\!\!\! \frac{4}{9}\ g_s^4(Q)\left(
2\frac{M_n^2}{\shat} - 2\frac{\tn\un}{\shat^2} + 1 \right) \, .
\end{eqnarray}

\noindent Again, for top production, the only change involves
correcting for the final state KK mass. The same results apply for
$\qt \qbt$ production.

For double KK quark production with different flavors in the final
state, the result is the same as the corresponding case with
identical flavors with the appropriate channel removed. That is,
$q q' \rightarrow \qs \qps$ is just the $t$-channel contribution
to $q q \rightarrow \qs \qs$,

\vspace{-8pt} \begin{eqnarray} \label{eq:dqqpqqp}
\mbox{\raisebox{-.5ex}{\Large$\bar{\Sigma}$}}\! \mid \!
\mathcal{M}(q q'  \rightarrow \qs \qps)\!\mid ^2 \, =& \!\!\!&
\!\!\!\!\! \frac{2}{9}\ g_s^4(Q)\left( -
M_n^2\frac{\shat}{\tn^2} + \frac{1}{4} + \frac{\shat^2}{\tn^2} \right) \, ,
\end{eqnarray}

\noindent while $q \bar{q}' \rightarrow \qs \qbps$ is also the
$t$-channel contribution to $q \bar{q} \rightarrow \qs \qbs$,

\vspace{-8pt} \begin{eqnarray} \label{eq:dqqbpqqbp}
\mbox{\raisebox{-.5ex}{\Large$\bar{\Sigma}$}}\! \mid \!
\mathcal{M}(q \bar{q}'  \rightarrow \qs \qbps)\!\mid ^2 \, =&
\!\!\!& \!\!\!\! \frac{1}{18}\ g_s^4(Q)\left( 4 M_n^2
\frac{\shat}{\tn^2} + 4 \frac{\shat^2}{\tn^2} + 8
\frac{\shat}{\tn} + 5 \right) \, ,
\end{eqnarray}

\noindent and similarly for \qt\ final states.

Finally, it is possible to produce the mixed KK final states
involving one \qs\ and one $\qt$. The projection operators
conspire to nullify the interference term in $q q \rightarrow \qs
\qt$. The differing signs of the $\gamma_5$'s also affect the $t$-
and $u$-channel contributions. The amplitude-squared for this
subprocess is:

\vspace{-8pt} \begin{eqnarray} \label{eq:dqqqsqt}
\mbox{\raisebox{-.5ex}{\Large$\bar{\Sigma}$}}\! \mid \!
\mathcal{M}(q q  \rightarrow \qs \qt)\!\mid ^2 \, =& \!\!\!&
\!\!\!\!\!\!\!\!\! \frac{1}{9} \ g_s^4(Q)\left[ 
\frac{2 M_n^2}{\shat} \frac{\shat^2}{\tn \un}
\left( \frac{\shat^2}{\tn \un} - 2 \right) \right.
\nonumber \\
&\!\!\! + &\!\!\!\! \left. 2 \frac{\shat^4}{\tn^2\un^2} - 8
\frac{\shat^2}{\tn\un} + 5 \right] \, .
\end{eqnarray}

\noindent The six remaining mixed subprocesses, $\qs \qbt$, $\qt
\qbs$, $\qs \qpt$, $\qt \qps$, $\qs \qbpt$, and $\qt \qbps$, all
are represented by the same $t$-channel diagram and have the same
form as the $t$-channel contribution to Eq.~\ref{eq:dqqqsqt}:

\vspace{-8pt} \begin{eqnarray} \label{e9}
\mbox{\raisebox{-.5ex}{\Large$\bar{\Sigma}$}}\! \mid \!
\mathcal{M}(q \bar{q}'  \rightarrow \qs \qbps)\!\mid ^2 \, =& &
\!\!\!\!\!\!\!\!\! \frac{1}{9}\ g_s^4(Q)\left[  2
\frac{M_n^2}{\tn}\left(1 + \frac{\un}{\tn}\right) 
+ \frac{5}{2} + 4 \frac{\un}{\tn}
+ 2 \frac{\un^2}{\tn^2}
\right] \, .
\end{eqnarray}

\noindent It is not possible to produce mixed KK final states from
initial gluons, nor is it possible to produce mixed KK final
states of a different flavor from initial $q \bar{q}$ pairs.

These amplitude-squared formulae do not contain any terms that
grow with energy, and the matrix elements for these subprocesses
are tree-unitary. This has also been observed for the case in
which only the gauge bosons propagate into the
bulk~\cite{gstar,dicus}. Note that the matrix elements of the
individual diagrams with external gluons are not tree-unitary:
There are delicate cancellations involved between individual
diagrams, which ensures unitarity for the total amplitude. As an
example, consider the subprocess, $q \bar{q} \rightarrow \gs \gs$,
which has both \qs\ and \qt\ propagators. The amplitude-squared
for this reaction would not be tree-unitary if there were just a
single tower of KK excitations of the quarks, or if the two towers
\qs\ and \qt\ did not couple left- and right-handedly to the SM
quarks. This is another example of tree-unitarity for a class of
massive vector boson theories other than the known spontaneously
broken gauge theories~\cite{unitary}.

\vspace{0.5cm}

\noindent {\bf 4. Stable KK Excitations}

\vspace{0.2cm}

\noindent As previously discussed, the lowest-lying KK excitations
of the light fermions and massless gauge fields may very well be
stable. This is a consequence of KK number conservation
(Eq.~\ref{eq:modes}), which is valid at all vertices and thus also
at the tree-level. KK number is broken at the loop-level, but the
lowest lying KK excitations of massless gauge bosons and the light
fermions can not decay even at the loop level\footnote{Loop corrections
may potentially create splitting between the masses of quark and massless
gauge boson KK excitations \cite{Cheng}, allowing, for example, for decays
such as $\gs \rightarrow q \qbs $ or $\qs \rightarrow q\gamma^\star$.
A short discussion of this case can be found in the next section.}
unless some new
physics mechanism is introduced. The KK excitations of massive
gauge bosons and heavier generation fermions can decay to lighter
KK states and SM fields at tree-level. For any SM decay with a
massless final state, such as $Z \rightarrow \nu \bar{\nu}$, there
are corresponding decays involving their KK excitations, such as
$Z_1^{\star} \rightarrow \nu_1^{\bullet} \bar{\nu}$. When the
final states are massive the decay may be kinematically forbidden,
depending on the compactification scale: For example, the
$t_1^{\bullet}$ can not decay to $W^{+} b_1^{\bullet}$ for a $400$
GeV compactification scale, but it can decay to $W_1^{+\star} b$.
At the tree-level, KK number conservation results in increasing
kinematic suppression of all decays involving KK excitations of
massive SM fields with increasing compactification scale. Note
also that the lowest-lying KK excitations of the quarks and gluons
can not decay to their SM counterparts via graviton emission
unless KK number is violated in such interactions. We consider the
hadronic collider phenomenology of stable or long-lived $n = 1$ KK
excitations in this section, then turn our attention to new
physics mechanisms that may result in short-lived lowest-lying KK
states and their associated phenomenology in the next section. By
long-lived, we refer to lifetimes long enough such that the final
state decay occurs beyond the detector.

For stable KK final states, the production cross sections for the
set of subprocesses $\{j\}$ enumerated in the previous section are
related to the squares of the amplitudes tabulated therein via:

\vspace{-8pt} \begin{eqnarray} \label{prod_crsec}
\sigma_{{}_{\mathit{KK}}}^{\mathit{tot}}
 =  \frac{1}{4\pi} & \!\!\!\!\! & \!\!\!\!\!\!\!
\sum_{j}\sum_{n} \,\,\,\,
\int_{\mbox{\raisebox{-1.5ex}{\scriptsize{$\!\!\!\!\!\!\!\!
\rho_n$}}}}^{\mbox{\raisebox{.9ex}{\scriptsize{$\!\!\!
1$}}}}dx_{{}_A}
\int_{\mbox{\raisebox{-1.5ex}{\scriptsize{$\!\!\!\!\!\!\!\!\!\!
\rho_n /x_{{}_A}$}}}}^{\mbox{\raisebox{.9ex}{\scriptsize{$\!\!\!
1$}}}}\!\!\!\!dx_{{}_B}
f_{a/{}_A}(x_{{}_A},Q) f_{b/{}_B}(x_{{}_B},Q) \nonumber \\
& \!\! & \!\!
\int_{\mbox{\raisebox{-1.3ex}{\scriptsize{$\!\!\!\!\!\!
-1$}}}}^{\mbox{\raisebox{.9ex}{\scriptsize{$\!\!\! 1$}}}} d z \,
\frac{\mbox{\raisebox{-.5ex}{\Large$\bar{\Sigma}$}}\! \mid \! M_j
\! \mid ^2}{S!} \, \frac{1}{ \hat{s} }\sqrt{1-\frac{4 M_n^2}{
\hat{s} }} \, ,
\end{eqnarray}

\noindent where $S$ is a statistical factor (the number of
identical final states) and $\rho_n = 4 M_n^2 / s$. The first
summation is over the subprocesses $\{j\}$ tabulated in the
previous section, while the second summation runs over all $n$ for
which pairs of final states with mass $M_n$ can be produced for a
given collider energy $\sqrt{s}$. The higher ($n
> 1$) states produce only a slight effect (at the $1\%$ level) due
to their large
mass.\addtocounter{footnote}{-1}\footnote{Furthermore, $Q = m_n$
for the $n > 1$ modes exceeds the compactification scale $\mu$,
for which the running of $\alpha_{{}_S} (Q)$ transforms from a
logarithmic to a power law behavior~\cite{unify}.  This has the
effect of reducing the contributions of the higher order
modes~\cite{topview} to the total cross sections even further.}
The cross sections for the higher modes are easily computed from
the cross section expression for the first mode by simply
replacing the mass of the first mode with that of the higher mode,
which includes adjusting the scale $Q$ to correspond to the higher
mass.

\begin{figure}
\centerline{ \includegraphics[height = 6.5in,width=5.in]{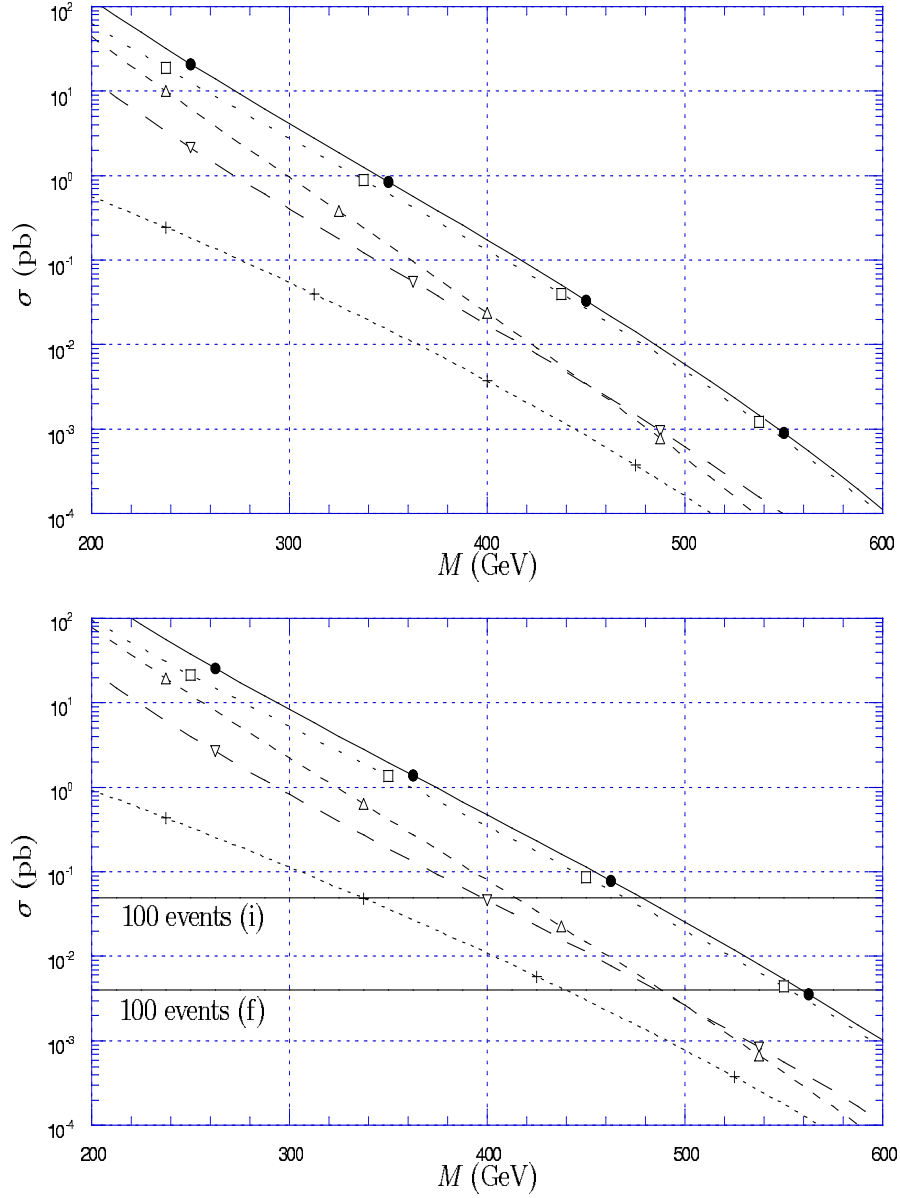}}
\caption{{The cross section for the production
of two stable KK final states is shown as a function of the KK
mass for Tevatron Run I (top) and II (bottom). The solid curve
corresponds to the total contribution, while the dashed lines
represent the partial contributions of KK quark pair ($\square$),
KK quark-gluon ($\vartriangle$), and KK gluon pair
($\triangledown$) production. Also shown is top production ($+$),
which features a different collider signature (namely, the top
will subsequently decay into additional states). Solid horizontal
lines mark $100$ events at the initial and final projected
luminosities for Run II.}} 
\label{fig:fTev}
\end{figure}

We evaluate the cross sections in Eq.~\ref{prod_crsec} with the
CTEQ$5$ distribution functions~\cite{CTEQ} and $Q = M_n$ in the
parton luminosity. In Fig.~\ref{fig:fTev},
we present the cross section for the production of two stable KK
final states for a given first excited KK mass $M = \mu = 1/R$ at
the Tevatron proton-antiproton collider. In addition to the total
cross section, the contributions of KK gluon pair, KK quark-gluon,
and KK quark pair production are plotted. For the case of double
KK quark production, the final state consists of light quark KK
excitations, but not the top quark, which can decay
({\textit{e.g.}}, $t_1^{\circ} \rightarrow W_1^{+\star} b$). The
production of KK quark pairs is dominant (not as much because the
cross section for a specific process is much higher, but because
there are many more processes involved), while the KK gluon pair
and KK quark-gluon production rates are comparable.

Stable, slowly moving KK quarks produced at colliders will
hadronize, producing high-ionization tracks. The production of
numerous heavy, charged stable particles will produce a clear
signal of new physics. They will appear as a heavy replica of the
light SM quarks, with both up- and down-type quark charges, but
with two KK quarks corresponding to each SM quark.

At the Tevatron Run I, searches for heavy stable quarks
\cite{connoly} have set an upper limit of about $1$ pb on the
production cross section of such particles (for a mass range
between $200$ and $250$ GeV). Using a naive extrapolation of the
limits presented in Ref.~\cite{connoly} to higher mass values, we
estimate a lower bound on the first excited KK mass of about $350$
GeV (in agreement with Ref.~\cite{ACD}). For the projected initial
(final) Run II ($\sqrt{s} = 2$ TeV) integrated luminosity, which
will yield $2$ ($25$) events  for each $10^{-3}$ pb of cross
section, $100$ events would be produced for a compactification
scale of $450$ GeV ($550$ GeV). In order to set definite limits on
the mass of KK excitations at Run II, an analysis similar to the
one performed for Run I is needed. An estimate of the Run II reach
can be made by assuming that the limit on the heavy stable quarks
production cross-section is driven by statistics. In this case, we
can expect an improvement of around a factor of $10$ in this
limit, to $0.1$ pb. Then, the nonobservation of heavy stable
quarks will raise the lower bound for the mass of the first KK
mode in the universal scenario to around $450$ GeV.

\begin{figure} 
\centerline{\includegraphics[height = 3.in,width=5.in]{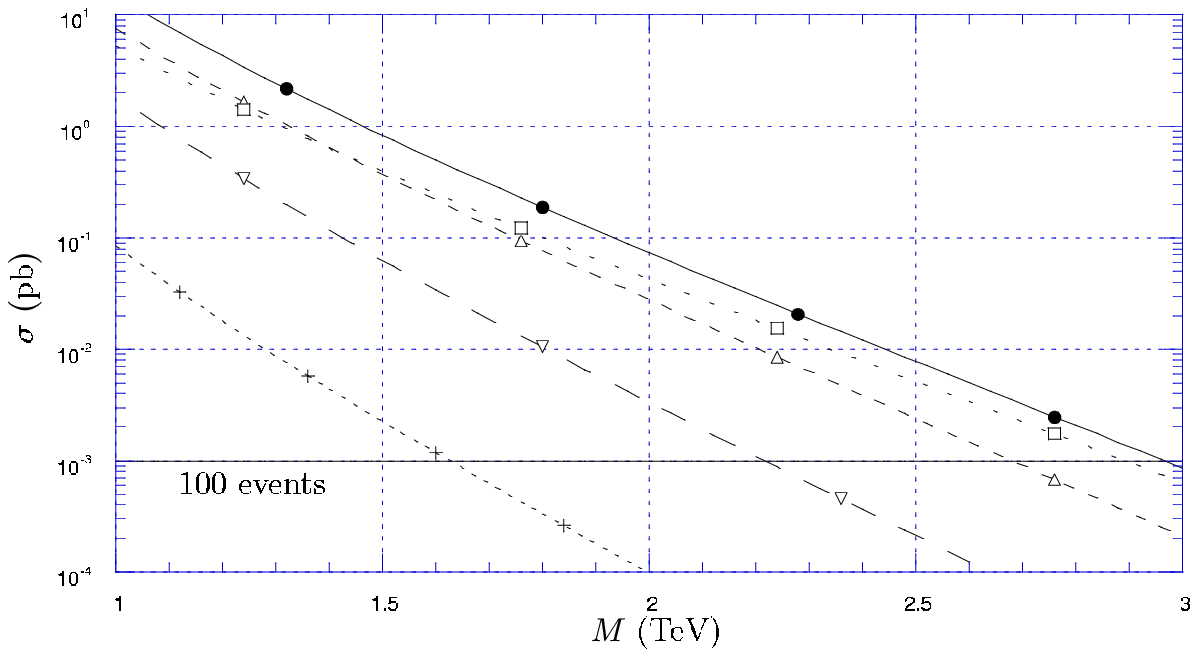}}
\vspace{40pt} \caption{\small{The same as Fig.~\ref{fig:fTev}, but
for the LHC. The solid horizontal line represents $100$ annual
events at the projected luminosity.}} 
\label{fig:fLHC}
\end{figure}

Much better prospects for the discovery of KK fields may be found
at the LHC proton-proton collider, where the anticipated annual
luminosity is $10^5$ pb$^{-1}$. The cross-sections for the
production rate of two stable KK excitations at the LHC energy are
illustrated in Fig.~\ref{fig:fLHC}.
A dedicated study is required to find the exact reach of the LHC
in this case, but, by requiring at least $100$ events to be
produced, we can estimate that the LHC will discover the first
stable KK excitations if their mass is smaller than about $3$ TeV.

Thus, stable KK quarks and gluons of the UED scenario will either
be discovered at the Tevatron Run II or the LHC, or  the lower
bound on their masses will be raised to around $450$ GeV or $3$
TeV, respectively. However, cosmological constraints require new
physics to explain the existence of stable KK excitations in this
mass range. This cosmological restriction can be lifted via a new
physics mechanism that causes the lowest-lying KK excitations to
have a lifetime that is short compared to the cosmological scale.
We now focus on this possibility.

\vspace{0.5cm}

\noindent {\bf 5. Decay Mechanisms}

\vspace{0.2cm}

\noindent The lowest-lying KK excitations of the light fermions
and the massless gauge bosons can decay into SM fields via new
physics mechanisms that produce a violation in KK number
conservation. Various decay schemes have been considered in the
literature \cite{Rujula}, \cite{ACD}, \cite{Rizzo}. However,
provided that the KK excitations decay within the detector, the
effect of a specific decay mechanism on the final state
distributions presented here can be expected to be small.

For purposes of illustration, we shall analyze in some detail the
decay properties of KK excitations in the fat brane scenario
proposed in Ref.~\cite{Rujula}. In this scenario, the ``small''
universal extra dimension is assumed to be the thickness of the
D$_4$ brane in which the SM particles propagate. In turn, this
brane is embedded in a $4 + N$ dimensional space, in which gravity
propagates. (In order to avoid drastically modifying Newton's law
at the solar system scale, we require $N \ge 2$.) We take the
gravity extra dimensions (call them $\{z_i\}$) to be symmetric,
with a compactification radius $r$  much larger than the thickness
of the fat brane $R$. The orbifold structure of the UED space in
which the SM fields propagate can be imposed by using boundary
conditions on the fat brane. The non-gravitational interactions
are identical to those presented in the Appendix. The differences
in this model lie in the interactions between gravity and the KK
excitations of the SM fields, where KK number violation in such
interactions will mediate the decays. The thick brane absorbs the
unbalanced momentum that results from the KK number violation.

The effective $4$D interactions of the graviton fields with the SM
fields and their KK excitations are obtained by the `naive'
(straightforward) generalization of the results in
Ref.~\cite{HLZ}. The Feynman rules for the couplings of the
graviton fields to the UED fields are related to the corresponding
couplings of the graviton fields to the SM fields by the form
factor $\mathcal{F}_n(x_y)$ as introduced in
Ref.~\cite{Rujula,Rizzo}. For example, the $\qs$-$q$-$G_{\vec{k}}$
coupling is:

\begin{equation} \Lambda_{\qs \mbox{-}q\mbox{-}G_{\vec{k}}} =
\mathcal{F}_n(x_y) \Lambda_{q\mbox{-}q\mbox{-}G_{\vec{k}}} \, ,
\end{equation}

\noindent where $G_{\vec{k}}$ is the KK excitation of the graviton
corresponding to mode $\vec{k}$ and $x_y \equiv m_y R = 2 \pi k_y
R / r$. Note that $n$ is the mode of the KK quark field, while
$k_y$ is the mode of the KK graviton field along the $y$
direction. Thus, $m_y$ is the contribution of the $y$ dimension to
the graviton mass. As with the non-gravitational interactions, the
KK quark field components associated with odd $Z_2$ parity
($Q_R(x)$, $U_L(x)$, and $D_L(x)$) do not interact with the SM
quark fields because of the presence of the projection operators.
Thus, these KK fields associated with odd $Z_2$ parity can not
decay to SM quarks and gravitons as indicated in
Ref.~\cite{Rizzo}. The form factor, $\mathcal{F}_n (x)$, does not
include the sine terms, and depends on the component of the
graviton mass arising from the universal compact dimension only,
$k_y$:

\begin{equation} \mathcal{F}_n(x_y) = \frac{\sqrt{2}}{\pi R}
\int_0^{\pi R} dy \exp(\frac{i 2 \pi k_y y}{r}) \cos(\frac{n
y}{R}) \, .
\end{equation}

\noindent Our result for the modulus-square of form factor,
\begin{equation} |\mathcal{F}_1 (x_y)|^2 =
\frac{4}{\pi^2}\frac{x_y^2}{(1-x_y^2)^2}\left[1+\cos(\pi
x_y)\right]
\end{equation}
\noindent differs by the sign of the cosine term from the one in
Ref.~\cite{Rizzo}, which is potentially significant, since it
affects the leading behavior of the form factor in the critical
regions: $x_y$ near zero (decay to light gravitons) and unity
(decay to heavy gravitons).

The total decay width is obtained by summing over all possible
graviton towers the partial decay width $\Gamma_n (x_y,x_z)$,
where $x_a$ refers to all of the extra dimensions, $x_y$ denotes
the universal direction, and $x_z$ is exclusive to gravity: $x_a^2
= x_z^2 + x_y^2$. The form-factor appears as a multiplicative
constant in the partial width:
\begin{equation} \Gamma_n (x_y,x_z) = |\mathcal{F}_n^2(x_y)|\ \Gamma'_n
(x_a) \, . \end{equation}
 Replacing the KK  sum with an integral over
the density of graviton states \cite{HLZ}, we obtain:

\begin{equation} \label{eq:Gtot} \Gamma_{\mathit{tot}} = \sum_{G,\Phi}
 \frac{ 2 \pi^{\frac{N-1}{2}}\
M_P^2\ M^{N}}{\Gamma(\frac{N-1}{2})\ \ M_D^{N+2}} \int_{2 \pi
R/r}^1 dx_y \ |\mathcal{F}_n^2(x_y)|\ \int_0^{\sqrt{1-x_y^2}}
x_z^{N-2} dx_z\ \Gamma'_n(x_a) \, .
\end{equation}

\noindent Here, $M_P$ is the conventional $4$D Planck scale, while
$M_D$ is the $(4 + N)$-dimensional Planck scale and should not be
more than one or two orders of magnitude above $1/R$ \cite{unify}.
Note that $N$ is the number of extra compact dimensions seen by
the graviton, as opposed to the number of universal dimensions,
which we take to be one.

For completeness, we give here the partial decay widths appearing
in Eq.~\ref{eq:Gtot}. These results are based on the three-point
vertex Feynman rules given in Ref.~\cite{HLZ}, with the masses of
all particles (except gravitons) set to
zero.\addtocounter{footnote}{-1}\footnote{This does not mean that
we neglect the KK mass of the particle decaying. Rather, this is a
consequence of the fact that the mass terms in the Feynman rules
in Ref.~\cite{HLZ} come from  mass terms in the Lagrangian that
are absent in the 5-dimensional theory.} The decay of the \qs\ (or
\qt\ ) into a SM quark and a massive spin 2 graviton $G^a$ has
partial width, apart from the overall form factor, given by:

\begin{equation} \label{eq:GqqG} \Gamma'{}_n (\qs
\rightarrow q G^a) = \frac{\kappa^2}{768 \pi} \frac{M_n^3}{x_a^4}
\left[ \left( 1 - x_a^2 \right)^4 \left( 2 + 3 x_a^2 \right)
\right] \, .
\end{equation}

\noindent The \qs\ can also decay into one of $N (N-1) /2$ massive
spin-$0$ particles, $\phi^a_{ij}$:

\begin{equation} \label{eq:GqqF} \Gamma'_n(\qs
\rightarrow q \phi^a_{ij}) = \delta_{ij} \frac{9 \kappa^2
\omega^2}{256 \pi} M_n^3 (1 - x_a^2)^2 \, ,
\end{equation}

\noindent where $\omega = \sqrt{\frac{2}{3(N+2)}}$. Finally, the
\gs\ can only decay into a SM gluon via massive spin 2 graviton
emission:

\begin{equation} \label{eq:GggG} \Gamma'_n(\gs
\rightarrow g G^a) = \frac{\kappa^2}{96 \pi} \frac{M_n^3}{x_a^4}
\left[ \left( 1 - x_a^2 \right)^2 \left( 1 + 3 x_a^2 + 6 x_a^4
\right) \right] \, .
\end{equation}

\begin{figure}
\centerline{\includegraphics[height = 3.in,width=5.in]{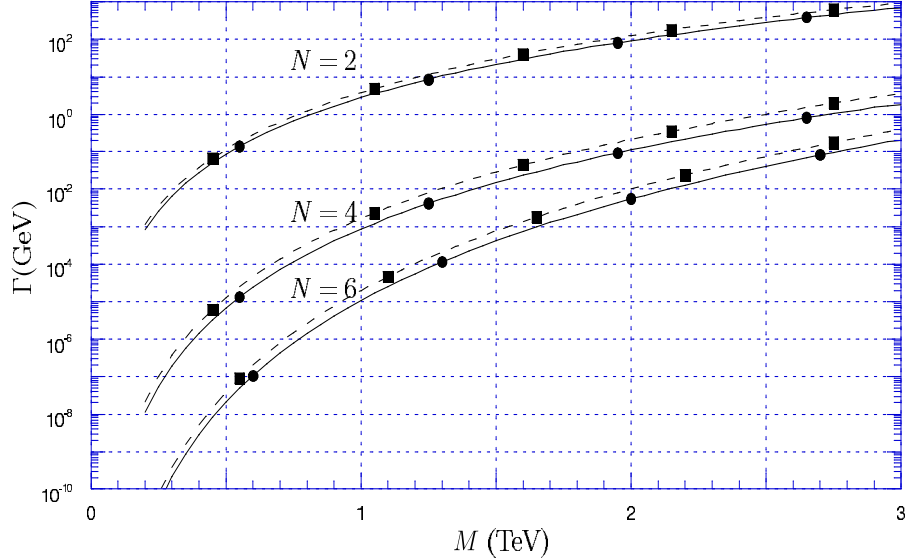}}
\vspace{0pt} \caption{\small{The decays of the $q_1^{\bullet}$ or
$q_1^{\circ}$ (solid) and $g_1^{\star}$ (dashed) into SM fields
via graviton emission (spin $2$ and scalar  combined) are shown as
a function of the compactification scale $\mu = M = 1/R$ for $M_D
= 5$ TeV. The pairs of curves correspond to $2$, $4$, and $6$
extra dimensions from top to bottom, respectively.}}
\label{fig:fDecay} 
\end{figure}

\begin{figure}
\centerline{\includegraphics[height = 6.5in,width=5.in]{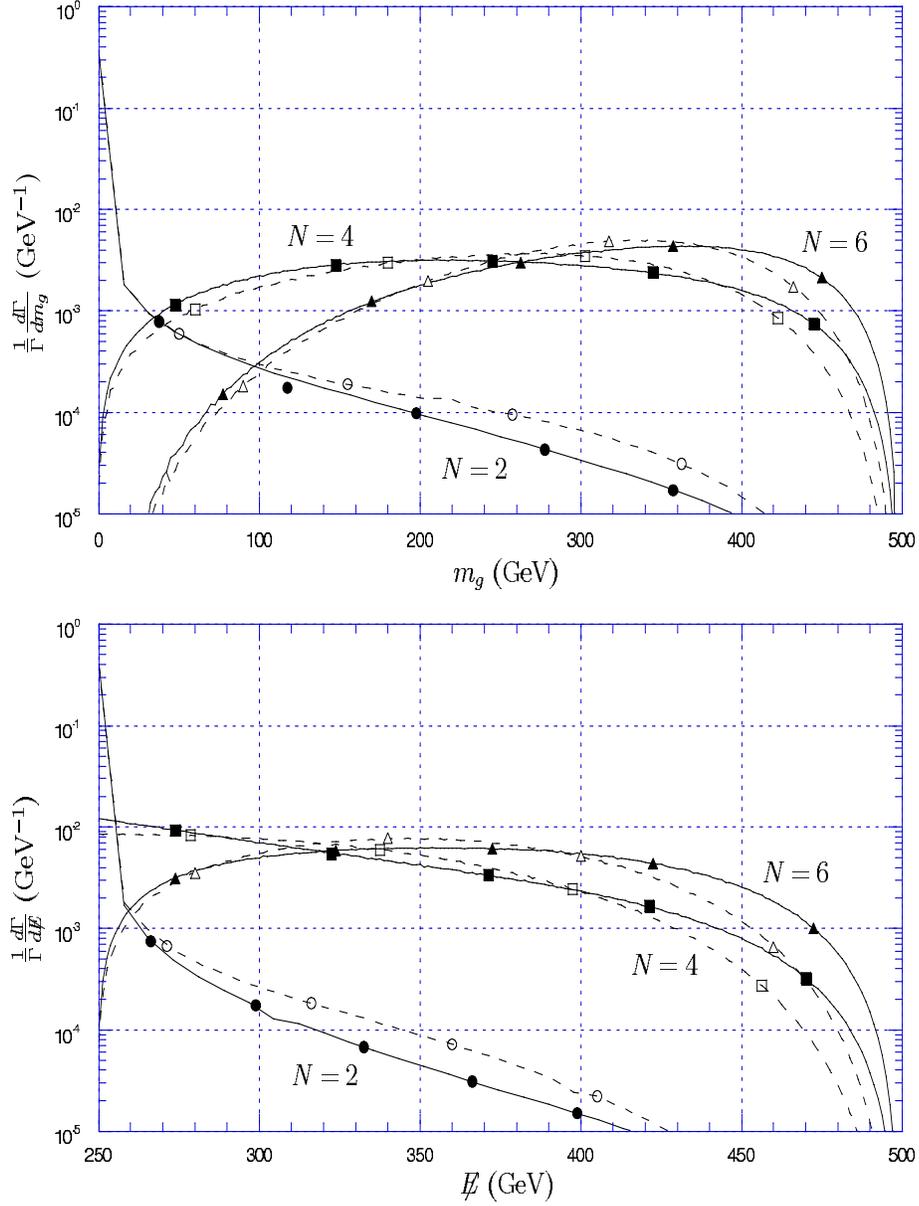} }
\caption{\small{The graviton mass distribution (top) and missing
energy distribution (bottom) of the $q_1^{\bullet}$ or
$q_1^{\circ}$ (solid) and $g_1^{\star}$ (dashed) are illustrated
for $\mu = 500$ GeV and $M_D = 5 TeV$. The pairs of curves
correspond to $2$, $4$, and $6$ extra dimensions.}}
\label{fig:mass_dist}
\end{figure}

The decay widths of the $q_1^{\bullet}$ (or $q_1^{\circ}$) and
$g_1^{\star}$, integrated over the density of graviton states with
the form factor as in the prescription of Eq.~\ref{eq:Gtot}, are
illustrated in Fig.~\ref{fig:fDecay}.
The distributions of the graviton mass and missing energy
(graviton energy) in the rest frame of the decaying particle are
shown in Fig.~\ref{fig:mass_dist}.
It is interesting to note that, in this scenario, when gravity
propagates in two extra-dimensions ($N = 2$), the decays of KK
quark or gluon excitations will be mediated mostly by very light
gravitons, while for $N \ge 3$ the heavy graviton (mass of order
$\mu$) contribution will dominate (see the top of
Fig.~\ref{fig:mass_dist}). As a consequence, for $N = 2$ the
missing energy distribution will have a peak at half the KK
excitation mass, while with increasing $N$ the distribution will
shift toward larger values. Note also that all of these decays
will occur within the detectors in the range of parameter space
that we will explore and is depicted here.

The collider signature for the production and decay of gluon or
light quark (except the top) KK excitations in this model is SM
dijet production with missing energy carried off by the gravitons.
This production rate is related to the cross sections for the
stable case and the differential branching fractions of the
decaying KK states via:

\begin{eqnarray}
d\sigma_{\mathit{tot}} = \sum_{A,B} d \sigma_{prod}( p \bar{p}
\rightarrow A\ B)\ \frac{ d \Gamma_A}{ \Gamma_A} \ \frac{ d
\Gamma_B}{ \Gamma_B} .
\end{eqnarray}

\noindent The sum is over the KK intermediate states, denoted by
$A$ and $B$. The spin correlations are not taken into account. The
top case will be discussed separately.


\begin{figure}
\centerline{\includegraphics[height = 6.5in,width=5.in]{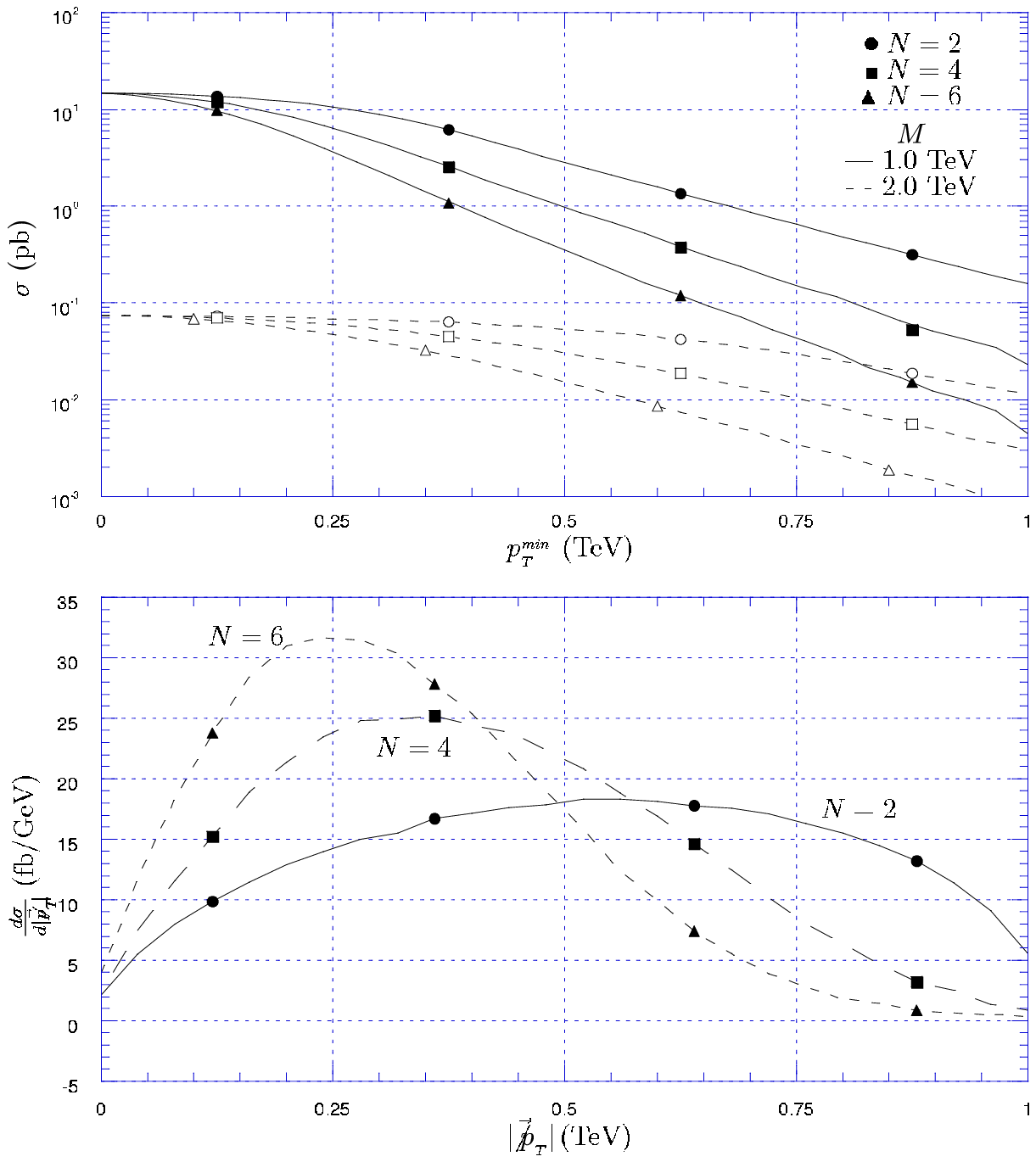} }
\caption{\small{The total cross section for the dijet production
plus missing energy from decaying KK final states (top) and the
missing transverse momentum
$\!\mid\!\!\vec{\not{p}}_{{}_T}\!\!\mid\!$ distribution (bottom)
are shown for $2$, $4$, and $6$ extra dimensions. The
compactification scale is $1$ TeV in the bottom figure, while $1$
and $2$ TeV are shown in the top figure. No cuts are implemented
in these graphs, such that the total area under each curve is
equal in the bottom graph. (However, all cuts are implemented in
the following figures.)}} \label{fig:pt_min}
\end{figure}

We consider the following two distributions of experimental
interest in Fig.~\ref{fig:pt_min}: the two-jets + missing energy
cross-section as a function of the minimum transverse momentum,
$p_{{}_T}^{{}_{\mathit{min}}}$, of the jets (top), and the
cross-section as a function of the missing transverse momentum,
$\!\mid\!\!\vec{\not{p}}_{{}_T}\!\!\mid\!$ (bottom).
The dependence of these distributions on the number of extra
dimensions in which gravity propagates (or on the decay mechanism)
is encoded in the mass distributions of the gravitons which
mediate this decay. For example, if the quark (or gluon) KK
excitations decay mostly to light gravitons, the distributions
will look like the curves corresponding to $N = 2$ in
Fig.~\ref{fig:pt_min}. Conversely, in the case when the KK
particles decay to heavy gravitons, these will take almost all
available momentum, leaving very little for the two observable
jets. Hence, the cross section drops faster with increasing
minimum transverse momentum, $p_{{}_T}^{{}_{\mathit{min}}}$, and
the missing transverse momentum,
$\!\mid\!\!\vec{\not{p}}_{{}_T}\!\!\mid\!$, distribution shifts
toward zero with the increase in $N$. Signals for decays mediated
by a different mechanism will fit somewhere among these curves,
depending on what fraction of the decays favor light versus heavy
gravitons.

\begin{figure} 
\centerline{\includegraphics[height = 6.5in,width=5.in]{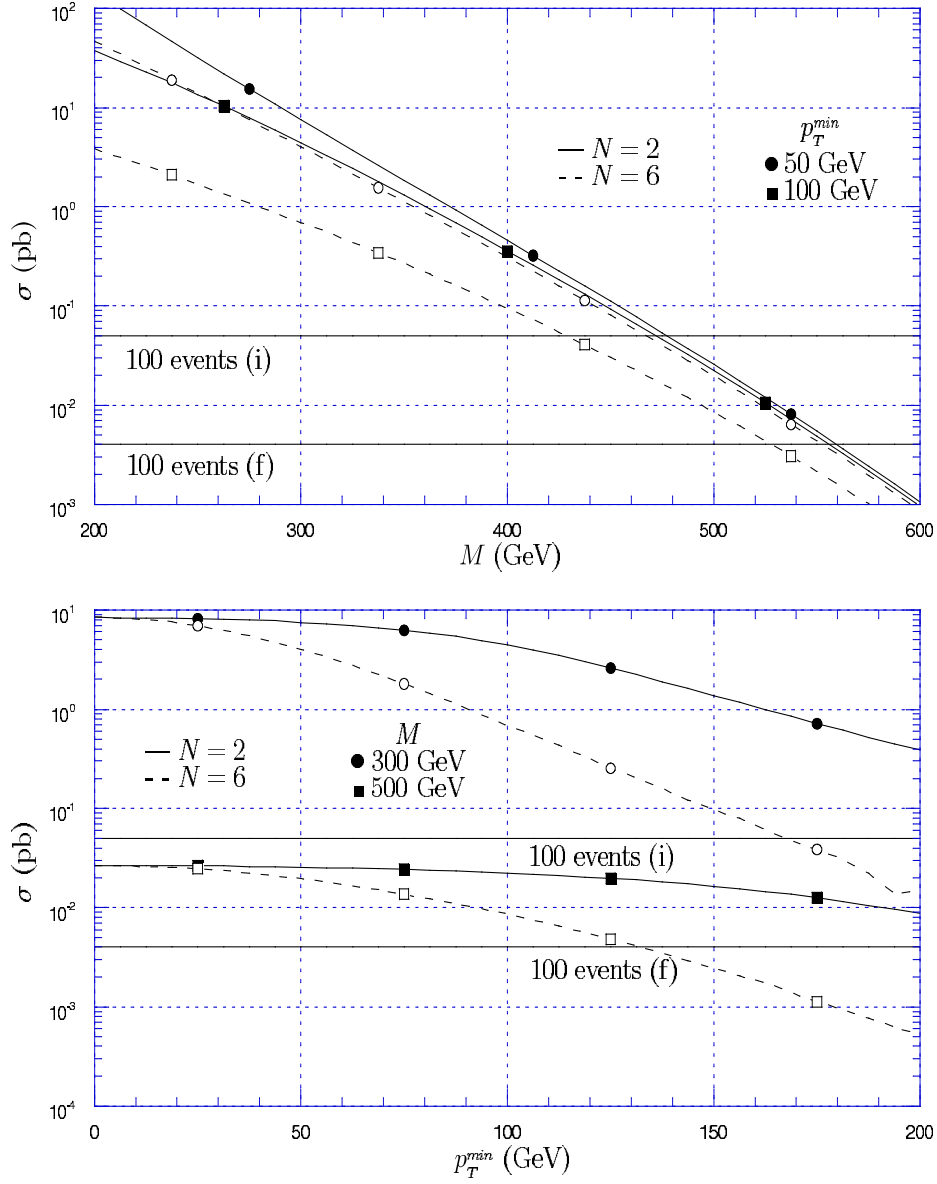}}
\vspace{0pt} \caption{\small{The total cross section for the dijet
production plus missing energy from decaying KK final states at
the Tevatron Run II energy is illustrated as a function of $\mu$
for fixed $p_{{}_T}^{{}_{\mathit{min}}}$ (top) and as a function
of the minimum transverse momentum $p_{{}_T}^{{}_{\mathit{min}}}$
for fixed values of the compactification scale $\mu$ (bottom).
Solid horizontal lines mark $100$ events at the initial and final
projected luminosities. In this and the following figures, we
implement cuts on the $p_{{}_T}$, rapidity, and separation of the
jets.}} \label{fig:fDTev}
\end{figure}

\begin{figure}
\centerline{\includegraphics[height = 6.5in,width=5.in]{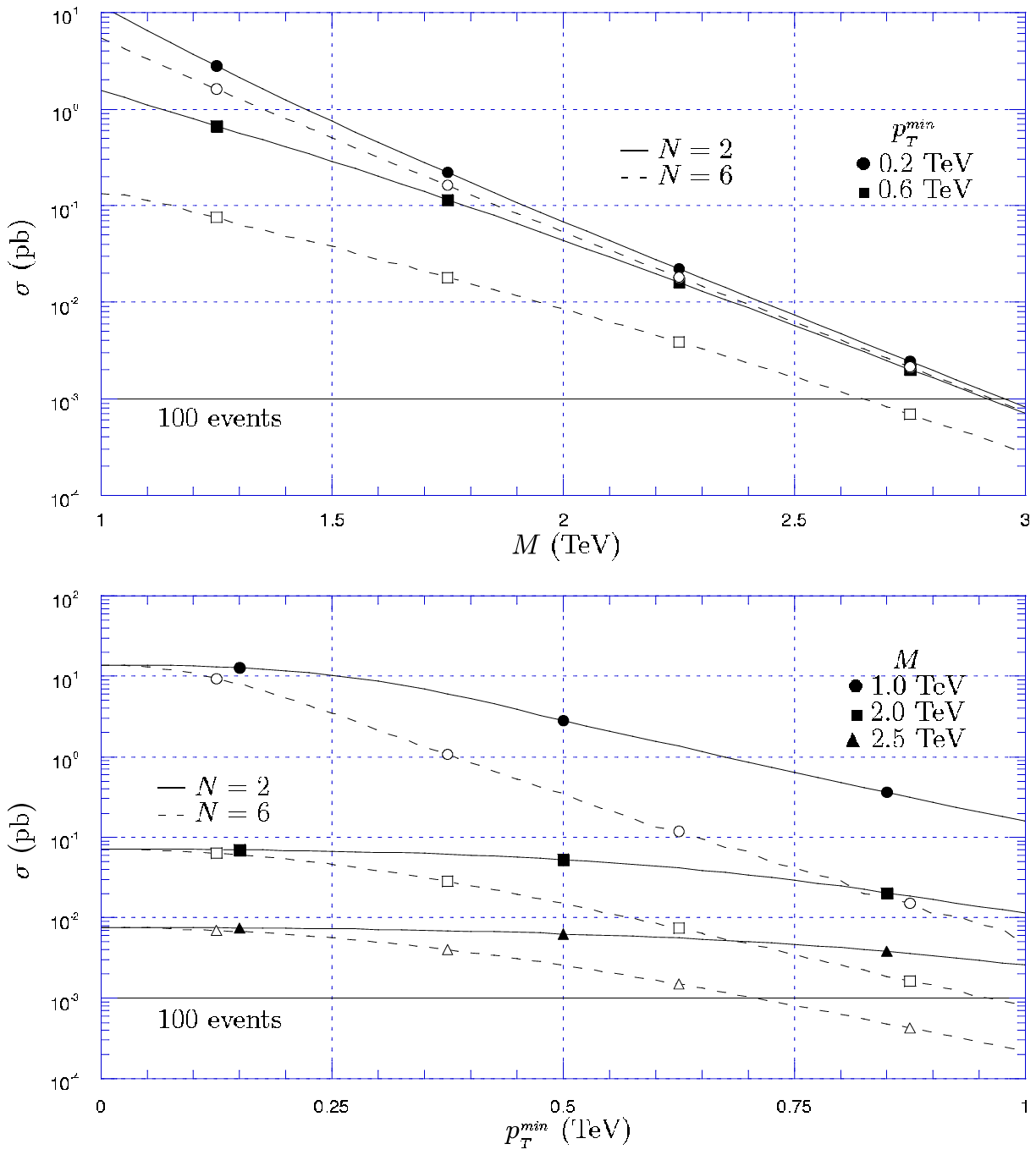}}
\vspace{0pt} \caption{\small{The same as Fig.~\ref{fig:fDTev}, but
for the LHC. The solid horizontal line marks $100$ annual events
at the projected luminosity.}}
\label{fig:fDLHC}
\end{figure}

\begin{figure} \setlength{\abovecaptionskip}{0pt}
\centering{\includegraphics[height = 6.5in,width=5.in]{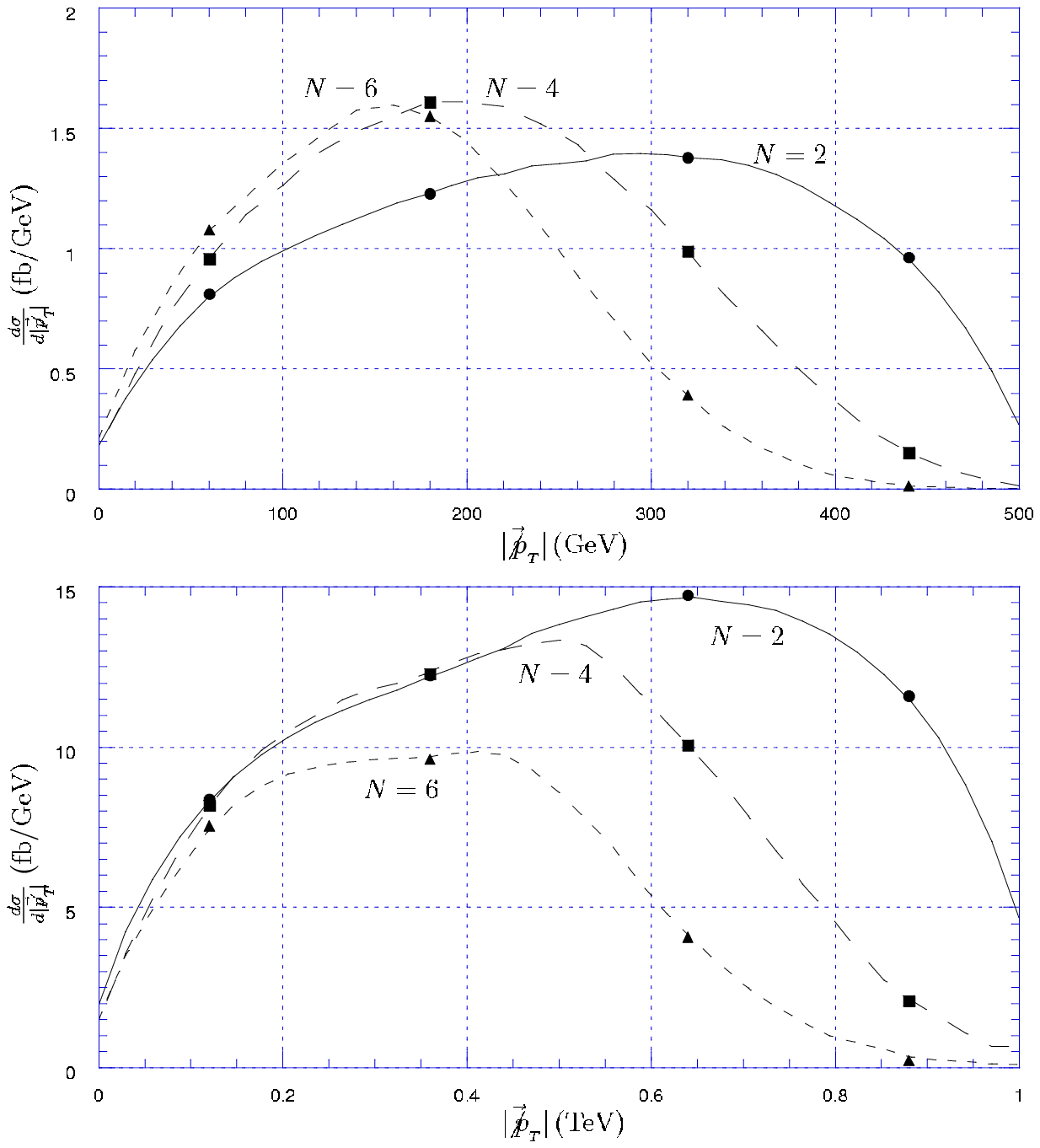}}
\vspace{0pt} \caption{\small{The missing transverse momentum
distribution is illustrated for Run II of the Tevatron (top) and
the LHC (bottom). The three curves represent $2$, $4$, and $6$
extra dimensions. By $\!\mid\!\!\vec{\not{p}}_{{}_T}\!\!\mid\!$ we
denote the vectorial sum of the transverse momentum of the two
emitted gravitons (which is equal and opposite to that of the
quarks). The compactification scale and minimum transverse
momentum are $400$ GeV and $50$ GeV for the Tevatron and $1$ TeV
and $200$ GeV for the LHC, respectively.}}
\label{fig:fdsdp}\setlength{\abovecaptionskip}{0pt}
\end{figure}

The dependence of the cross section on the mass of the KK
excitations for different $p_{{}_T}$ cuts is shown in
Fig.~\ref{fig:fDTev} for the Tevatron Run II and
Fig.~\ref{fig:fDLHC} for the LHC.
For illustration, the values of $N = 2$ and $N = 6$ for the number
of extra dimensions have been used. Note that the case $N = 6$ is
the least favorable to direct observation, since the heavier the
graviton mass, the lower the transverse momentum of the quark or
gluon jets will be. Beside the cuts specified in the figure, we
also require that the rapidity be limited to the range $\mid\! y
\!\mid \, \leq 2.5$, and the two observable jets be separated by a
cone of radius larger than $R = \sqrt{(\Delta\phi)^2 +
(\Delta\eta)^2} = 0.4$, where $\phi$ is the azimuthal angle and
$\eta$ is the pseudorapidity, which is related to the polar angle
$\theta$ via $\eta = - \ln \tan (\theta/2)$. Requiring for direct
observation at least $100$ events with $p_{{}_T} > 50$ GeV at
Tevatron and $p_{{}_T} > 400$ GeV at LHC, respectively, we see
that the Tevatron reach extends to about $550$ GeV, while at LHC
KK excitations can be discovered in this model for values of the
compactification scale as high as $3$ TeV. We assume here that
cuts on missing transverse momentum (Fig.~\ref{fig:fdsdp}) are
used to greatly reduce the SM background.

We present here some comments on the SM background. There are many SM
processes which can give rise to a dijet  signal with missing energy.
Some examples include $WZ,\ ZZ,\ q \bar{q} Z$ and $t \bar{t}$ production,
where neutrinos arising from $Z$ and $W$, for example, carry off the mising
energy; also $2 \rightarrow 2$ QCD processes with missing energy due to the 
mismeasurement of jet energies. Of course, cuts on the minimum $p_T$ of the
jets and on the missing transverse energy can be implemented to greatly 
improve the signal-to-background ratio. A complete analysis of SM
backgrounds (including the optimization of cuts) is beyond the purpose of 
this paper. However, for illustration, we consider the specific cuts in Table II.
For example, for $p_T^{min} = 600 $ GeV and $|\vec{\not{p}}_T| > 1200 $GeV
at the LHC, the SM background has been evaluated in Ref. \cite{Bityukov} to
be $\sim 40$ events for $10^5$ pb$^{-1}$ luminosity, while the signal would be 
600, 2000, and 50 events for $N=2$ and compactification scale $M = 1, 2,$ and 3 TeV
respectively. For $N=6$, the signal would be 30, 130 and 10 events, for
the same values of $M$. We see that the signal is larger than or comparable 
with the background in almost all of these cases ($N=6, M= 3$ TeV is
borderline). Moreover, these cuts can be optimized in order to
enhance the signal-to-background ratio. For example, in the case of
$M = 1$ TeV, the 1200 GeV cut on the missing transverse energy is too hard
(this is why so few events remain), and by relaxing it the signal can be increased 
substantially.

Finally, we consider the production and decay of KK excitations of
the top quark. As seen from Fig. \ref{fig:fTev}, the cross-section
for this process is less than $1\%$ than the total KK excitation
production cross-section. However, if the mass of first KK tower
is smaller than about $1$ TeV, there will be of order $10^4$ KK
top pair events produced at LHC.
Unlike the other KK excitations, the $t^{\bullet}$ can also decay
to $W^{+\star} b$. For $\mu < 1$ TeV, the decay to $W^{+\star} b$
is dominant (unless $N = 2$; in this case, we need $\mu < 0.4$
TeV). Furthermore, the $W^{+\star}$ can decay either into $W$ +
graviton, in which case the signal for this process will be
$b\bar{b}W^+W^-$ in the final state, plus missing energy; or into
$d^{\bullet} u$, for example, in which case the signal could be
two $b$ jets plus four light quark jets plus missing energy.

The results discussed thus far apply to the case when the first KK
excitations of quarks and gluons have nearly the same masses. This is
true at tree level; however, radiative corrections can lift this mass
degeneracy \cite{Cheng}. In this situation, the decays of the first 
KK excitations can proceed through cascades to the lightest KK particle
(LKP). For example, if the LKP is the $\gamma^\star$ (as in \cite{Cheng}),
the $\gs$ can decay through $\gs \rightarrow q \qbs \rightarrow  q \qb \gamma^\star$.
The case when the LKP is stable has been analyzed in Ref. \cite{CMSexp} and
the collider phenomenology has been found to be very similar to that
of supersymmetry with an almost degenerate spectrum. Here we want to
comment on the possibility that the LKP decays through a gravity mediated
mechanism, as discussed above in this section. In this case, the collider 
signal will be an excess of two photon events instead of two jets (or two
leptons, if the LKP is an $l^\bullet$, for example). Moreover, since the momenta
of the SM particles radiated in the process of cascade decays to the LKP
(the two quarks in the $\gs$ decay example above) should be rather small
(of the order of the mass splitting between the different KK excitations),
the momentum of the LKP wil be nearly the same as the momentum of the KK
particle initiating the decay (the $\qs$ or $\gs$). Then the $p_T$ and missing 
energy distrinutions of the two photon (or two lepton) events will be the
same as the distributions computed above for the dijet case. We leave a
more complete analysis (including branching fractions for gravity mediated
decays versus cascade decays to the LKP) to a future paper \cite{cmn2}. 

\begin{table}[!t]
\begin{center}
\begin{tabular}{|c| c| c c |c c |c c c|}
\hline
\hline
 & & \multicolumn{6}{c}{Signal (evts.)} &\\
$p_T^{cut}$ & \hbox{Background} &   \multicolumn{2}{c}{$M =1$ TeV} 
& \multicolumn{2}{c}{$M =2$ TeV} & \multicolumn{2}{c}{$M =3$ TeV}&\\
(GeV)& (evts.)& $N = 2$ & $N = 6$
& $N = 2$ & $N = 6$ & $N = 2$ & $N = 6$ &\\
\hline
100 & $3\times 10^6$ & $1\times 10^6$ & $9\times 10^5$ 
& $7\times 10^3$ & $6\times 10^3$ & $84$ & $80$ &\\
200 & $2\times 10^5$ & $9\times 10^5$ & $2\times 10^5$ 
& $6\times 10^3$ & $4\times 10^3$ & $80$ & $65$ &\\
300 & $9\times 10^3$ & $4\times 10^5$ & $4\times 10^4$ 
& $5\times 10^3$ & $3\times 10^3$ & $73$ & $50$ &\\
400 & $1\times 10^3$ & $1\times 10^5$ & $2\times 10^3$ 
& $4\times 10^3$ & $1\times 10^3$ & $65$ & $34$ &\\
500 & $2\times 10^2$ & $5\times 10^4$ & $2\times 10^2$ 
& $3\times 10^3$ & $4\times 10^2$ & $58$ & $20$ &\\
600 & $4\times 10\  $ & $6\times 10^2$ & $3\times 10\ $ 
& $2\times 10^3$ & $1\times 10^2$ & $50$ & $10$ &\\
\hline
\hline
\end{tabular}
\end{center}
\label{table1}
\caption{SM background and UED signals with $p_T > p_T^{cut}, \
\not{E_T}> 2 p_T^{cut}$ (for $10^5 \hbox{pb}^{-1}$ at LHC). }
\end{table}

\vspace{0.5cm}

\noindent {\bf 6. Conclusions}

\vspace{0.2cm}

\noindent In this work, we have investigated in detail the
phenomenology of the UED model, which is a class a class of
string-inspired models in which all of the SM fields can propagate
into one TeV-scale extra dimension. Specifically, we calculated
the effects that the KK excitations of the quarks and gluons have
on multijet final states at high energy hadronic colliders
including the LHC and Tevatron Runs I and II. We performed these
calculations for the case where the lowest-lying KK excitations of
the light quarks and gluons are stable, as well as the case where
they decay within the detector. For the decaying scenario, we
examined a scenario in the context of a fat brane that may provide
enough KK number violation to accommodate lifetimes that would be
consistent with cosmological observations without resulting in a
significant production rate for single KK final states. We
presented a detailed evaluation for the fat brane scenario, and
also illustrated the dependence of our results on the decay
structure.

Our results for proton-proton collisions at the Tevatron Run I
place the mass bound for the first excited KK states at
$350$--$400$ GeV. For the Run II energies, the mass bound can be
raised to $450$--$550$ GeV. Proton-antiproton collisions at the
LHC energy can probe much further: UED KK excitations will either
be discovered or the mass limit will be raised to about $3$ TeV.
If the UED compactification scale is less than $1.5$ TeV, then at
the LHC energy we might be able to see the first two KK
excitations of the quarks and gluons, thereby uniquely
establishing the extra-dimensional nature of the new physics.

The signatures of the production of UED KK excitations will be
vastly different for short-lived and long-lived states. Stable,
slowly moving KK quarks produced at colliders will hadronize,
resulting in tracks with high ionization. The production of
numerous heavy, charged stable particles will produce a clear
signal of new physics. They will appear as a heavy replica of the
light SM quarks, with both up- and down-type quark charges, but
with two KK quarks corresponding to each SM quark. The two towers,
\qs and $\qt$, will be polarized with opposite chirality for all
cross-channel processes due to $Z_2$ parity conservation. If the
KK excitations of the light quarks and gluons are short-lived,
then the signal will be SM dijet production with missing energy
carried off by the emitted gravitons. This missing energy
significantly reduces the SM background. The production of the
lowest-lying KK excitations of the gluons and light quarks gives
rise to only dijets plus missing energy (due to the escaping
gravitons), and no multijet signals (at order $\alpha_S^2$). Such
final states will distinguish this new physics from supersymmetry,
which will produce multijet final states in addition to dijets.

\vspace{0.5cm}

\noindent {\bf Acknowledgments}

\vspace{0.2cm}

\noindent We are grateful to X. Tata for
useful discussions. 
This work was supported in part by the U.S. Department
of Energy Grant Numbers DE-FG03-98ER41076 and DE-FG02-01ER45684.

\vspace{0.5cm}

\noindent {\bf Appendix
\appendix}

\vspace{0.2cm}

\noindent We begin with the UED $5$D Lagrangian density. The
procedure for obtaining the effective $4$D theory is to Fourier
expand the $5$D fields in terms of the extra dimension $y$, and
then integrate over $y$. Here, we will begin by obtaining the mass
contributions to the KK excitations from their kinetic terms as
well as their interactions with the Higgs potential. We will then
proceed to derive the complete set of interactions between the KK
excitations of the quarks and gluons. We will not discuss purely
gluonic interactions, which were described elaborately in
Ref.~\cite{gstar}.

Each of the $5$D multiplets $Q(x,y)$, $U(x,y)$, and $D(x,y)$ can
be Fourier expanded in terms of the compactified dimension $y$,
restricted in an $S_1 / Z_2$ orbifold, as

\vspace{-9pt} \begin{eqnarray} Q (x,y) &\!\!\! = &\!\!\!
\frac{1}{\sqrt{\pi R}} \left\{ \left( \! \begin{array}{c} u (x)
\\ d (x) \end{array} \! \right)_{\!L} + \sqrt{2} \sum_{n=1}^{\infty}
\left[ Q_L^n (x) \cos \left(\frac{n y}{R} \right) + Q_R^n (x) \sin
\left(\frac{n y}{R} \right) \right] \right\} \label{eq:Qdecomp} \\
U (x,y) &\!\!\! = &\!\!\! \frac{1}{\sqrt{\pi R}} \left\{ u_R (x) +
\sqrt{2} \sum_{n=1}^{\infty} \left[ U_R^n (x) \cos \left(\frac{n
y}{R} \right) + U_L^n (x) \sin \left(\frac{n y}{R} \right) \right]
\right\} \label{eq:Udecomp} \\
D (x,y) &\!\!\! = &\!\!\! \frac{1}{\sqrt{\pi R}} \left\{ d_R (x) +
\sqrt{2} \sum_{n=1}^{\infty} \left[ D_R^n (x) \cos \left(\frac{n
y}{R} \right) + D_L^n (x) \sin \left(\frac{n y}{R} \right) \right]
\right\} \label{eq:Ddecomp} \, ,
\end{eqnarray}

\noindent where $Q_{L,R}^n (x) \equiv \frac{1}{2} (1 \mp \gamma_5
) \qs (x)$ as in Eq.~\ref{eq:id} and $\gamma_5$ is the usual $4$D
Dirac matrix. Note that the decomposition in Eq.'s
\ref{eq:Qdecomp}--\ref{eq:Ddecomp} gives the correct SM zero mode
chiral structure for the fermions. Similarly, the gluon field $A_M
(x,y)$ can be Fourier expanded as:

\vspace{-9pt} \begin{eqnarray} A_{\mu}^a (x,y) =
\frac{1}{\sqrt{\pi R}}\left[ A_{\mu 0}^a (x) +
\sqrt{2} \sum_{n=1}^{\infty}A_{\mu ,n}^{a} (x) \cos(\frac{n y}{R}) \right] \\
A_{4}^a (x,y) = \frac{\sqrt{2}}{\sqrt{\pi R}}
\sum_{n=1}^{\infty}A_{4,n}^{a} (x) \sin(\frac{n y}{R}) \, .
\end{eqnarray}

\noindent Under the transformation $y\rightarrow -y$, the
decomposed gluon fields transform as $A_{\mu}^a (x,-y) = A_{\mu}^a
(x,y)$ and $A_4^a (x,-y) = - A_4^a (x,y)$. Notice that $Z_2$
parity and KK number are conserved in the interactions involving
the gauge fields and fermions. We choose to work in the unitary
gauge, where we can apply the gauge choice $A_{4,n}^a (x) = 0$
\cite{gauge}.

The primary contribution to the KK masses stems from the kinetic
term in the Lagrangian density:

\vspace{-9pt} \begin{equation} \label{eq:LkinQ2} \mathcal{L}_5  =
i \bar{Q} (x,y) \left\{ \Gamma^M \left[ \partial_M + i \g T^a
A_M^a (x,y) \right] \right\} Q (x,y) \, .
\end{equation}

\noindent There are similar terms for the other $5$D multiplets.
Here, \g\ is the $5$D strong coupling and $M$ is the $5$D analog
of the Lorentz index $\mu$, \textit{i.e.}, $M \in \{\mu,4\}$.
Integration of the kinetic terms in Eq.~(\ref{eq:LkinQ2}) over the
compactified dimension $y$ results in:

\vspace{-9pt} \begin{eqnarray} i \!\!
\int_{\mbox{\raisebox{-1.3ex}{\scriptsize{$\!\!\!\!
0$}}}}^{\mbox{\raisebox{.9ex}{\scriptsize{$\!\!\!\! \pi R$}}}} &
\!\!\!\!\!\!\!\! \bar{Q} (x,y) & \!\!\!\! \Gamma^M
\partial_M Q (x,y) dy
= i \left[ (\bar{u}(x) \bar{d}(x))_L \gamma^{\mu}
\partial_{\mu} \left( \!
\begin{array}{c} u (x)
\\ d (x) \end{array} \! \right)_{\!L} \right. \nonumber \\
&\!\!\! + &\!\!\!\!\!\!\! \sum_{n=1}^{\infty} \left.
\bar{Q}_L^n(x) \gamma^{\mu}
\partial_{\mu} Q_L^n(x) + \bar{Q}_R^n(x) \gamma^{\mu}
\partial_{\mu} Q_R^n(x) \right. \\
&\!\!\! + &\!\!\!\!\!\!\! \left. i\frac{n}{R}\bar{Q}_L^n (x)
Q_R^n(x) + i\frac{n}{R}\bar{Q}_R^n (x) Q_L^n(x) \right] \nonumber
\, .
\end{eqnarray}

\noindent There are similar expressions for the $U(x,y)$ and
$D(x,y)$ multiplets. The mass of the KK excitations are identified
as $n \mu$, where $\mu$ is the compactification scale ($\mu =
1/R$). Thus, in the absence of the Higgs mechanism, the KK
excitations have masses given by $M_n = n/R = n \mu$. The
corresponding mass matrix is:

\vspace{-9pt} \begin{equation} (\bar{Q}^n (x) , \bar{U}^n (x) )
\left(
\begin{array}{cr} \frac{n}{R} & 0 \\ 0 & -\frac{n}{R}
\end{array}\right) \left( \! \begin{array}{c} Q^n (x)
\\ U^n (x) \end{array} \! \right) \, , \nonumber \end{equation}

\noindent where $Q^n (x)$ represents the upper component of the
doublet, with charge $2/3$. Note that there is no mixing between
the different KK levels, {\textit i.e.}, between $Q^n (x)$ and
$Q^m (x)$ for $n \neq m$.

Additional mass contributions arise from the Yukawa couplings of
the $5$D quark multiplets via the Higgs VEV's:

\vspace{-9pt} \begin{eqnarray} & \!\!\!\!i& \!\!\!\!
\int_{\mbox{\raisebox{-1.3ex}{\scriptsize{$\!\!\!\!
0$}}}}^{\mbox{\raisebox{.9ex}{\scriptsize{$\!\!\!\! \pi R$}}}}
\left[ \lambda_u^5 \bar{Q} (x,y) i \sigma_2 H^{\ast}(x,y) U(x,y) +
\lambda^5_d \bar{Q}(x,y) H(x,y) D(x,y) + h.c. \right] dy =
\nonumber \\
& \!\!\!\!i& \!\!\!\! \left\{ M_u \left[ \bar{u} (x) u(x) +
\sum_{n=1}^{\infty} \left[ \bar{Q}_L^n(x) U_R^n(x) +
\bar{Q}_R^n(x) U_L^n(x) \right] \right] \right. \\
& \!\!\!\! + & \!\!\!\! \left. \lambda_u \left[ \bar{u} (x) u(x)
h(x) + \sum_{n=1}^{\infty} \left[ \bar{Q}_L^n(x) U_R^n(x) +
\bar{Q}_R^n(x) U_L^n(x) \right] h(x) \right] + \lambda_d
\textnormal{terms} \right\} \nonumber \, ,
\end{eqnarray}

\noindent where $\lambda_u \equiv \lambda_u^5 / \sqrt{\pi R}$ and
$M_u \equiv \lambda_u \!\! <\!\!H\!\!>$. The $(Q^n(x),U^n(x))$
mass matrix, including these Yukawa contributions as well as the
kinetic terms, is:

\vspace{-9pt} \begin{equation} (\bar{Q}^n (x) , \bar{U}^n (x) )
\left(
\begin{array}{cr} \frac{n}{R} & M_u \\ M_u & -\frac{n}{R}
\end{array}\right) \left( \! \begin{array}{c} Q^n (x)
\\ U^n (x) \end{array} \! \right) \, . \nonumber \end{equation}

The eigenvalues of the this mass matrix give the net mass $M_n$ of
the KK modes in terms of the mass of the corresponding quark field
$M_q$ and the mass from the compactification $n/R$:

\vspace{-9pt} \begin{equation} M_n = \sqrt{\frac{n^2}{R^2} +
M_q^2} \, .
\end{equation}

\noindent We redefine the $U^n(x)$ field by $U^n(x) \rightarrow
\gamma_5 U^n(x)$. In our subsequent calculations, we neglect the
SM quark masses except for the top mass $M_t$.

The interactions between the $5$D $Q(x,y)$ fields and the $5$D
gluon fields $A_M^a(x,y)$ are given by:

\vspace{-9pt} \begin{eqnarray} -&\!\!\!\!\!g_{{}_5} &\!\!\!\!\!\!
\int_{\mbox{\raisebox{-1.3ex}{\scriptsize{$\!\!\!\!
0$}}}}^{\mbox{\raisebox{.9ex}{\scriptsize{$\!\!\!\! \pi R$}}}}
\bar{Q} (x,y) \Gamma^M T^a A_M^a (x,y) Q (x,y) d y \nonumber \\
= -& \!\!\!\!\!\!\! g &\!\!\!\!\! \left.
\mbox{\raisebox{-.5ex}{\Huge$\{$}} \bar{q}_L (x) \gamma^{\mu} T^a
q_L (x) A_{\mu ,0}^{a} (x) + \sum_{n=1}^{\infty} \left[\bar{Q}_L^n
(x) \gamma^{\mu} T^a Q_L^n(x)
+ \bar{Q}_R^n \gamma^{\mu} T^a Q_R^n (x) \right] A_{\mu ,0}^{a}(x) \right. \nonumber \\
& \!\!\! + & \!\!\! \sum_{n=1}^{\infty} \left[ \bar{q}_L(x)
\gamma^{\mu} T^a Q_L^n(x) + \bar{Q}_L^n(x) \gamma^{\mu} T^a q_L(x)
\right]
A_{\mu ,n}^{a} (x) \\
& \!\!\! + & \!\!\! \frac{1}{\sqrt{2}} \sum_{n,m,\ell =1}^{\infty}
\left. \left[ \bar{Q}_L^n (x) \gamma^{\mu} T^a Q_L^m(x)
(\delta_{\ell,\mid m - n \mid}+\delta_{\ell,m+n}) \right. \right. \nonumber \\
& \!\!\! + & \!\!\! \left. \left. \bar{Q}_R^n (x) \gamma^{\mu} T^a
Q_R^m (x) (\delta_{\ell,\mid m - n \mid}-\delta_{\ell,m+n})
\right] A_{\mu ,\ell}^{a} \mbox{\raisebox{-.5ex}{\Huge$\}$}}
\right. \nonumber \, ,
\end{eqnarray}

\noindent where $g \equiv \g / \sqrt{\pi R}$. There are similar
interactions involving the $U$ and $D$ fields. In terms of the
\qs\ and \qt\ fields (Eq.~\ref{eq:id}), the interactions are:

\vspace{-9pt} \begin{eqnarray} \mathcal{L}_{\mathit{int}} = &
\!\!\! -&\!\!\!\!\! g \mbox{\raisebox{-.5ex}{\Huge$\{$}}\!\!
\left. \bar{q} (x) \gamma^{\mu} T^a q (x) A_{\mu ,0}^{a} (x) +
\sum_{n=1}^{\infty} \left[ \qbs (x) \gamma^{\mu} T^a \qs (x) + \qt
(x) \gamma^{\mu} T^a \qbt (x) \right]
 A_{\mu ,0}^{a}(x) \right. \nonumber \\
& \!\!\! + & \!\!\! \sum_{n=1}^{\infty} \left[ \bar{q}_L(x)
\gamma^{\mu} T^a \qs (x) + \qbs (x) \gamma^{\mu} T^a q_L(x)
\right] A{}_{\nu ,n}^{a} (x) \nonumber \\
& \!\!\! + & \!\!\! \sum_{n=1}^{\infty} \left[ \bar{q}_R(x)
\gamma^{\mu} T^a \qt (x) + \qbt (x) \gamma^{\mu} T^a q_R(x)\right]
A{}_{\nu ,n}^{a} (x) \\
& \!\!\! + & \!\!\! \frac{1}{\sqrt{2}} \sum_{n,m,\ell =1}^{\infty}
\left[ - \qbs (x) \gamma^{\mu} \gamma_5 T^a \qs[m] (x) + \qbt (x)
\gamma^{\mu} \gamma_5 T^a \qt[m] (x) \right] A_{\mu ,\ell}^{a} \,
\delta_{\ell, m + n} \nonumber \\
& \!\!\! + & \!\!\! \frac{1}{\sqrt{2}}\sum_{n,m,\ell =1}^{\infty}
\left. \left[ \qbs (x) \gamma^{\mu} T^a \qs[m] (x) + \qbt (x)
\gamma^{\mu} T^a \qt[m] (x) \right] A_{\mu ,\ell}^{a} \,
\delta_{\ell, \mid m - n \mid} \right.
\!\!\mbox{\raisebox{-.5ex}{\Huge$\}$}} \nonumber \, .
\end{eqnarray}

\noindent The relative coupling strengths are summarized in
Fig.~\ref{fig:feynq}.

\vspace{0.5cm}

\bibliographystyle{unsrt}

\begin{thebibliography}{99}

\bibitem{Planck} N. Arkani-Hamed, S. Dimopoulos and G. Dvali,
Phys. Lett. B {\bf 429}, 263 (1998); Phys. Rev. D{\bf 59}, 086004
(1999); I. Antoniadis, N. Arkani-Hamed, S. Dimopoulos and G.
Dvali, Phys. Lett. B {\bf 436}, 257 (1998).

\bibitem{superstring} E. Witten, Nucl. Phys. B {\bf 471}, 135
(1996); J. Lykken, Phys. Rev. D {\bf 54}, 3693 (1996).

\bibitem{collider} See, for example:  E.A. Mirabelli, M.
Perelstein, and M.E. Peskin, Phys. Rev. Lett. {\bf 82}, 2236
(1999); G.F. Giudice, R. Rattazzi, and J.D. Wells, Nucl. Phys.
B {\bf 554}, 3 (1999);  J.E. Hewett, Phys. Rev. Lett. {\bf
82}, 4765 (1999); G. Shiu and S.H.H. Tye, Phys. Rev. D{\bf 58},
106007 (1998); T. Banks, A. Nelson, and M. Dine, J. High Energy
Phys. {\bf 06}, 014 (1999); P. Mathews, S. Raychaudhuri, and S.
Sridhar, Phys. Lett. B{\bf 450}, 343 (1999) and J. high Energy Phys.
{\bf 07}, 008 (2000); T.G.
Rizzo, Phys. Rev. D {\bf 59}, 115010 (1999); C. Balazs, H.-J. He,
W.W. Repko, C.-P. Yan, and D.A. Dicus, Phys. Rev. Lett. {\bf 83},
2112 (1999); I. Antoniadis, K. Benakli, and M. Quir\'{o}s, Phys.
Lett. B {\bf 360}, 176 (1999); P. Nath, Y. Yamada, and M.
Yamaguchi, {\it ibid.} {\bf 466}, 100 (1999); W.J. Marciano, Phys.
Rev. D {\bf 60}, 09006 (1999); T. Han, D. Rainwater, and D.
Zepenfield, Phys. Lett. B {\bf 463}, 93 (1999); K. Aghase and N.G.
Deshpande, {\it ibid.} {\bf 456}, 60 (1999); G. Shiu, R. Shrock,
and S.H.H. Tye, {\it ibid.} {\bf 458}, 274 (1999); K. Cheung and
Y. Keung, Phys. Rev. D {\bf 60}, 112003 (1999); K.Y. Lee, S.C.
Park, H.S. Song, J.H. Song and C.H. Yu, Phys. Rev. D {\bf 61},
074005 (2000); hep-ph/0105326.

\bibitem{HLZ}T.~Han, J.~D.~Lykken and R.~J.~Zhang,
  Phys.\ Rev.\ D {\bf 59}, 105006 (1999).

\bibitem{asym} J. Lykken and S. Nandi, Phys. Lett. B {\bf 485}, 224 (2000).

\bibitem{astro} V.~Barger, T.~Han, C.~Kao and
R.J.~Zhang,
Phys.\ Lett.\  B {\bf 461}, 34 (1999); S. Cullen and M. Perelstein,
Phys. Rev. Lett. {\bf 83}, 268, 1999; L.J. Hall and D.~Smith,
Phys.\ Rev.\  D {\bf 60}, 085008 (1999); S.C. Kappadath et. al.,
Bull. Am. Astron. Soc. {\bf 30} 926 (1998); Ph.D. Thesis, available at
http://wwwgro.sr.unh.edu/users/ckappada/ckappada.html; D.A. Dicus,
W.W. Repko and V.L. Teplitz, Phys. Rev. D {\bf 62}, 076007 (2000).
M. Biesiada and B. Malec, {\it ibid.} {\bf 65}, 043008;
K.A. Milton, R. Kantowski, C. Kao
and Y. Wang, Mod. Phys. Lett. A {\bf 16}, 2281 (2001);
S. Hannestad, G. Raffelt, Phys. Rev. Lett. {\bf 87}, 051301 (2001);
M. Fairbarin, J. High Energy Phys. {\bf 02}, 024 (2002).


\bibitem{Antoniadis} I. Antoniadis, Phys. Lett. B {\bf 246} 377 (1990).

\bibitem{asymcoll} E. Accomando, I. Antoniadis, and K. Benakli,
Nucl. Phys. B {\bf 579}, 3 (2000); A. Datta, P.J. O'Donnell, Z.H.
Lin, X. Zhang, and T. Huang, Phys. Lett. B {\bf 483}, 203 (2000).

\bibitem{ew} T.G. Rizzo and J.D. Wells, Phys.\ Rev.\  D {\bf 61}, 016007 (2000) ;
C.D. Carone, Phys. Rev. D {\bf 61} 015008, 2000.

\bibitem{muon} M.~Masip and A.~Pomarol,
Phys.\ Rev.\  B {\bf 60}, 096005 (1999);
P.~Nath, Y.~Yamada and M.~Yamaguchi, Phys.\ Lett.\  B {\bf 466},
100 (1999).

\bibitem{estar} C.D. McMullen and S. Nandi,
hep-ph/0110275.

\bibitem{gstar} D.A. Dicus, C.D. McMullen and S. Nandi,
 Phys.\ Rev.\ D {\bf 65}, 076007 (2002).

\bibitem{ACD} T. Appelquist, H.-C. Cheng and B.A. Dobrescu,
Phys.\ Rev.\  D {\bf 64}, 035002 (2001).

\bibitem{Rujula} A. DeRujula, A. Donini, M.B. Gavela and S.
Rigolin, Phys. Lett. B {\bf 482}, 195 (2000).

\bibitem{Rizzo} T.G. Rizzo, Phys.\ Rev.\  D {\bf 64}, 095010
(2001).

\bibitem{UED} T. Appelquist and B.A. Dobrescu, Phys. Lett.
B {\bf 516}, 85 (2001); K. Agashe, N.G. Deshpande and G.H. Wu,
Phys. Lett. B {\bf 514}, 309 (2001).

\bibitem{rescale} A. Delgado, A. Pomarol and M. Quir\'{o}s,
Phys. Rev. D {\bf 60}, 095008 (1999); J. High Energy Phys. {\bf 01}, 030 (2000);
A. Pomarol and M. Quir\'{o}s, Phys. Lett. B {\bf 438}, 255 (1998);
 E. Dudas, Class. Quantum Grav. {\bf 17}, R41 (2001); M. Masip and A. Pomarol,
Phys. Rev. D {\bf 60}, 096005 (1999).

\bibitem{gauge} K.R. Dienes, E. Dudas, and T. Ghergetta, Nucl.\
Phys.\ B {\bf 537}, 47 (1999); J. Papavassiliou and A. Santamaria
Phys.\ Rev.\ D {\bf 63}, 125014 (2001).

\bibitem{FORM} J.A.M. Vermaseren, math-ph/0010025.

\bibitem{dicus} R.S. Chivukula, D.A. Dicus
and H.-J. He, Phys.\ Lett.\ B {\bf 525}, 175 (2002).

\bibitem{unitary} J.M. Cornwall, D.N. Levin and G. Tiktopoulos,
             Phys. Rev. Lett. {\bf 30}, 1268 (1973); Phys. Rev.
             D {\bf 10}, 1145 (1974); D.A. Dicus and V.S. Mathur,
             Phys. Rev. D {\bf 7}, 3111 (1973); B.W.Lee, C. Quigg
             and H.B.Thacker, Phys. Rev. Lett. {\bf 38}, 883 (1977);
             Phys. Rev. D {\bf 16}, 1519 (1977); M. J. G. Veltman,
             Acta Phys. Polon. B {\bf 8}, 475 (1977); C.H. Llewellyn Smith,
             Phys. Lett. B {\bf 46}, 233 (1973).

\bibitem{Cheng}
  H.~C.~Cheng, K.~T.~Matchev and M.~Schmaltz,
  Phys.\ Rev.\ D {\bf 66}, 036005 (2002).

\bibitem{unify} K.P. Dienes, E. Dudas and T. Gherghetta,
Phys. Lett. B{\bf 436}, 55 (1998); Nucl. Phys. B{\bf 537}, 47
(1999); T. Taylor and G Veneziano, Phys. Lett. B{\bf 212} 147
(1988); D. Ghilencia and G.G. Ross, Phys. Lett. B{\bf 442} 165
(1998); hep-ph/9908369; C. Carone, Phys. Lett. B{\bf 454} 70
(1999); P.~H.~Frampton and A.~Rasin,
Phys.\ Lett.\  B{\bf 460}, 313 (1999); A.~Delgado and
M.~Quir\'{o}s,
Nucl.\ Phys.\  B{\bf 559}, 235 (1999); A.~Perez-Lorenzana and
R.~N.~Mohapatra,
Nucl.\ Phys.\  B{\bf 559}, 255 (1999); Z.~Kakushadze and
T.~R.~Taylor,
Nucl.\ Phys.\  B{\bf 562}, 78 (1999); D. Dumitru and S. Nandi,
Phys. Rev. D {\bf 62}, 046006 (2000);
 K.~Huitu and T.~Kobayashi, Phys.\ Lett.\  B{\bf
470}, 90 (1999); H.~Cheng, B.~A.~Dobrescu and C.~T.~Hill,
Nucl.\ Phys.\  B{\bf 573}, 597 (2000).

\bibitem{topview}
M.~Masip,
  Phys.\ Rev.\ D {\bf 62}, 105012 (2000).

\bibitem{CTEQ} H.L. Lai \emph{et al.}, Phys.\ Rev.\ D {\bf
51}, 4763 (2000).

\bibitem{connoly} A. Connolly [CDF collaboration], ''Search
for long-lived charged massive particles at CDF,'' Talk at the
American Physical Society (APS) Meeting of the Division of
Particles and Fields (DPF $99$), Los Angeles, CA, Jan $5$-$9$,
$1999$, hep-ex/9904010.

\bibitem{Bityukov}
  S.~I.~Bityukov and N.~V.~Krasnikov,
  Phys.\ Lett.\ B {\bf 469}, 149 (1999).

\bibitem{CMSexp}
  H.C. Cheng, K.T. Matchev and M. Schmaltz,
  Phys. Rev. D {\bf 66}, 056006 (2002).

\bibitem{cmn2}
  C. Macesanu, C.D. McMullen and S. Nandi,
  Phys. Lett. B {\bf 546}, 253 (2002).

\bibitem{Webber}
  J.~M.~Smillie and B.~R.~Webber,
  JHEP {\bf 0510}, 069 (2005).

\end{thebibliography}

\end{document}